\documentclass[fleqn,10pt]{wlscirep}
\usepackage[utf8]{inputenc}
\usepackage[T1]{fontenc}
\usepackage{placeins}
\usepackage[mathlines]{lineno}
\usepackage{xcolor}
\usepackage{algpseudocode}
\usepackage{algorithm}

\title{Shock propagation from the Russia-Ukraine conflict on international multilayer food production network determines global food availability}

\author[1,2]{Moritz Laber}
\author[1,3,4]{Peter Klimek}
\author[5,6]{Martin Bruckner}
\author[1]{Liuhuaying Yang}
\author[1,3,4,7,*]{Stefan Thurner}
\affil[1]{Complexity Science Hub Vienna, A-1080 Vienna, Austria}
\affil[2]{Network Science Institute, Northeastern University, Boston, MA 02115, USA}
\affil[3]{Center for Medical Data Science CeDAS, Medical University of Vienna, A-1090 Vienna, Austria}
\affil[4]{Supply Chain Intelligence Institute Austria, A-1080 Vienna, Austria}
\affil[5]{Institute for Ecological Economics, Vienna University of Economics and Buisness, A-1020 Vienna, Austria}
\affil[6]{ETH Zurich, Institute of Environmental Engineering, 8093 Zurich, Switzerland}
\affil[7]{Santa Fe Institute, Santa Fe, NM 85701, USA}

\affil[*]{stefan.thurner@meduniwien.ac.at}

\begin{document}
\flushbottom
\maketitle

\thispagestyle{empty}
    

\begin{quote}
{\sffamily\small
\textbf{\large Abstract}\vskip2pt
Dependencies in the global food production network can lead to shortages in numerous regions, as demonstrated by the impacts of the Russia-Ukraine conflict on global food supplies. Here, we reveal the losses of $125$ food products after a localized shock to agricultural production in $192$ countries and territories using a multilayer network model of trade (direct) and conversion of food products (indirect), thereby quantifying $10^8$ shock transmissions. We find that a complete agricultural production loss in Ukraine has heterogeneous impacts on other countries, causing relative losses of up to $89\%$ in sunflower oil and $85\%$ in maize via direct effects, and up to $25\%$ in poultry meat via indirect impacts. Whilst previous studies often treated products in isolation and did not account for product conversion during production, our model studies the global propagation of local supply shocks along both production and trade relations, allowing comparison of different response strategies.}
\end{quote}
\vspace{10pt}

\section{Introduction}\label{sec1}

Trade relations among countries create a global network~\cite{Ercsey-Ravasz_2012-05, DOdorico_2014,MacDonald_2015-03, Dupas_2019-05,Fair_2017-08, BerfinKarakoc_2021 , DeRaymond_2021-06}. This trade network facilitates the propagation of locally confined shocks in the food system around the globe~\cite{Puma_2015-02,DAmour_2016-02,Kummu_2020-03, Gaupp_2020-06,DeRaymond_2021-06}. Such shocks can result from a variety of often overlapping causes, most notably extreme weather events or economic and geopolitical crises~\cite{Gephart_2017-01, Cottrell_2019-02,Reichstein_2021-04}, and they have been found to become more frequent over time~\cite{Cottrell_2019-02}. Building dynamical models of shock propagation~\cite{Tamea_2016-01,Gephart_2016-02,Marchand_2016-09,Burkholz_2019-10,Heslin_2020, Grassia_2022-03} makes it possible to assess the effect of a locally confined event on the economy in distant places and to compare different response strategies~\cite{Gephart_2016-02,Heslin_2020}. These modelling efforts have highlighted that the set of affected countries extends beyond direct trading partners~\cite{Burkholz_2019-10} and that countries differ in their ability to deal with shocks depending on their position in the trade network~\cite{Tamea_2016-01, Gephart_2016-02} next to their access to food reserves~\cite{Marchand_2016-09,Heslin_2020}.
Previous work~\cite{Distefano_2018-08} has shown that food crises are not always correlated with spiking food prices, thereby calling for methods that complement the modeling of food prices~\cite{Schewe_2017-04, Falkendal_2021-01}.\\
Even though it has been recognized that shocks co-occur in different parts of the food system~\cite{Cottrell_2019-02}, shock propagation models have so far often treated each commodity in isolation and neglected that products may be converted into other products along the food production chain. Input--output models provide a well-established formalism to account for the conversion of products into each other~\cite{Galbusera_2018-09,Miller_2021}. In the context of food systems, this framework was successfully employed to shed light on the use of resources in foreign countries~\cite{Bruckner_2019-10,Sun_2020-06}. As a demand-driven model the input--output formalism is, however, less suitable to assess the propagation of shocks caused by changes in supply rather than demand~\cite{Galbusera_2018-09,Pichler_2021-07}.
\\
Here, we present a multilayer network model that takes into account both trade between countries and the production dependencies among products. We base all model parameters on data on supply and use of individual food products in different countries~\cite{Bruckner_2019-10}. Our model allows us to simulate shocks to the production of individual products and to assess the resulting losses of the same and other products in countries around the globe.
We employ our model in three different case studies, that are based on the ongoing war in Ukraine, one of the worlds largest producers of maize, wheat, and sunflower seeds~\cite{FAO_2022-05}: First, we simulate a shock assuming a complete loss of agricultural production in Ukraine and show that the availability of various products in different world regions is severely reduced. Second, we study production shocks across the entire spectrum of food products in Ukraine and the resulting loss of different products. Here, we focus on maize and sunflower oil, which make up the largest shares of Ukraine's food exports. 
Finally, to demonstrate the versatility of our model to explore other kinds of food shocks, we consider the availability of pork in Germany and identify critical suppliers and production inputs, i.e. country and product pairs that would reduce the availability of pork in Germany if they suffer a shock.
With this work we therefore establish a tool to assess multiple impacts emerging from shocks on the interconnected trade and production networks.
\section{Results}\label{s:results}
\begin{figure*}[h]
    \centering
    \includegraphics[width=\textwidth]{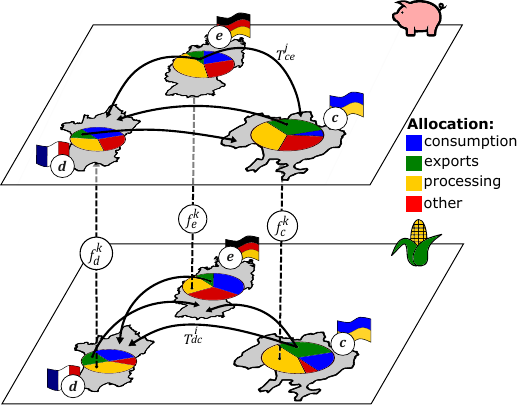}
    \vspace{0.4cm}
    \caption{Schematic representation of trade and production as a multilayer network for three countries, Ukraine, $c$, France, $d$, and Germany, $e$, and two products maize (lower layer $i$) and pigs (upper layer $j$). The allocation of products to different purposes is represented as a pie chart within each country. Trade is described by the weighted directed links (solid arrows) within each layer. The entry $T_{cd}^i$ describes the share of country $d$'s exports (green) of a product, $i$, directed towards country $c$. Production processes, modelled as a second type of node, turn products into other products, thereby connecting different layers (dashed arrows). Here, the production function, $f^k_c$, for the process type, $k$, pig husbandry, in country $c$, turns maize into pigs. Country $c$ is therefore an in-neighbor of the process on the maize-layer and an out-neighbor on the pig-layer. Note that production processes can take more than one input and supply more than one output.}
    \label{fig:model_sketch}
\end{figure*}

\paragraph{Understanding trade and production as a multilayer network}
Our model describes trade and production of different food products in different countries as an iterated three-step process. These steps are (i) the allocation of products to different purposes, (ii) trade with other countries and (iii) food conversion and processing activities, i.e. production of products using other products as input. Figure\,\ref{fig:model_sketch} represents a simplified version of our model illustrated by means of a multilayer network. For simplicity, it includes only two products, maize and pigs, and three countries, Ukraine, France and Germany. 
First, a country, $c$, possesses an amount, $x_c^i(t)$, of a product, $i$, in iteration $t$, and constitutes a node in the multilayer network. In a first step this amount is allocated in fixed but country- and product-specific proportions to  different purposes, namely consumption as food, export, further processing and other uses. This split is represented as a pie chart within each country in fig.\,\ref{fig:model_sketch}.\\
Second, countries trade with other countries. A country, $c$, directs a fraction, $T^i_{dc}$, of its exports of product $i$ towards country $d$. The trade of a product, $i$, corresponds to a single layer of the multilayer network and the fractions, $T^i_{dc}$, form a matrix describing this weighted, directed %
In a third step, countries produce new products by converting input products to output products using different types of production processes. The production process of type $k$ in a country, $c$, is modelled by a production function, $f^k_c$, which maps the available amount of input products to the amount of output products. These functions constitute a second type of node, that acts as an intermediary between replicas of the same country on different product-layers. In the example depicted in fig.\,\ref{fig:model_sketch} the process pig husbandry connects countries in the maize-layer to their replica in the pig-layer, as maize can be used to feed pigs. The simplified depiction does not show other possible fodder crops included in the model nor does it depict other processes that use maize as an input. \\
Performing each of the three steps once in every country constitutes a single iteration (model time step) of the algorithm. We define a baseline scenario that consists of $10$ iterations of the dynamics described above and denote $\underline{x}^i_c(t)$ the amount of a product, $i$, in a country, $c$ in iteration $t$. We obtain parameters and initial conditions, $\underline{x}^i_c(t=0)$, from trade and production data\cite{Bruckner_2019-10} of the year $2013$ (see methods). The trade matrices $T^i$ and production functions $f^k_c$ are not re-calibrated to data during the simulation but stay fixed. The variable $x^i_c(t)$ is updated according to the production, trade and allocation steps and, therefore, changes each iteration. As we calibrate our model parameters to yearly data, we can think of an iteration as a model year.

\paragraph{Shocks propagate through different channels}

\begin{figure*}[p]
    \centering
    \includegraphics[width=\textwidth]{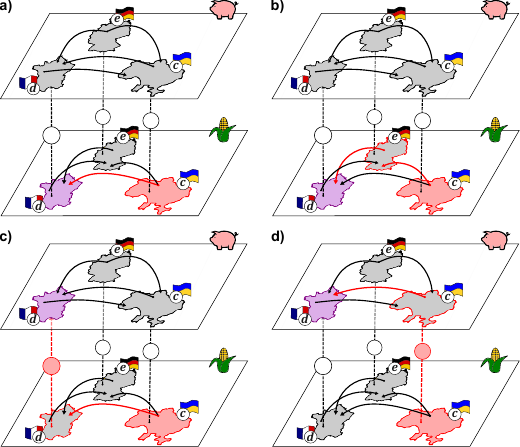}
    \vspace{0.4cm}
    \caption{A toy example that illustrates different shock propagation channels that can contribute to losses in our model. The shocks to the agricultural production of a good, in this example maize, in one country (red), here Ukraine, can propagate along different channels (red arrows) and thereby induce losses in another country (purple), here France.~\textbf{a)} A direct shock affects direct neighbors in the trade network of the shocked product.~\textbf{b)} Shock propagation on the trade network leads to indirect shocks that arise through trade via intermediaries, in this example Germany.~\textbf{c)} In our model not only the shocked product but also products, that rely on it as an input to be produced, are affected. In this case, France lacks a foreign input, Ukrainian maize, to feed pigs.~\textbf{d)} The shocked product is also not available for production in the shocked country, thereby reducing the exports of those products that rely on it as an input. This opens up another shock propagation channel. In the example, Ukraine produces less pigs as it lacks maize as fodder and therefore exports less pigs to France.}
    \label{fig:shock}
\end{figure*}

We compare the baseline scenario to a shocked scenario. In the shocked scenario, the produced amount of one or more products in a specific country is removed from the model in each time step. We denote the amount of a product, $i$, in a country, $c$, after a shock to product $j$ in country $d$ as $X^{ji}_{dc}(t)$, where $t$ denotes the time step. The relative loss of a product, $i$, in a country, $c$, after a shock to a product, $j$, in a country, $d$,

\begin{equation}\label{eq:RelativeLoss}
    \mathrm{RL}^{ji}_{dc}(t) = \frac{\underline{x}^i_c(t) - X^{ji}_{dc}(t)}{\underline{x}_c^i(t)}~,
\end{equation}

describes the relative reduction of the amount of product $i$ in country $c$, in the shocked scenario with respect to the baseline scenario at iteration $t$. We omit the dependence on the iteration $t$ when referring to the relative loss in the last iteration $t_\mathrm{end}$, $\mathrm{RL}^{ji}_{dc}=\mathrm{RL}^{ji}_{dc}(t_\mathrm{end})$.\\
The quantity $\mathrm{RL}$ captures the combined effect of several shock propagation channels. We illustrated the different shock propagation channels with a toy example in fig.\,\ref{fig:shock}. The direct trade relations of a country comprise only one such channel (fig\,\ref{fig:shock}a)). The propagation of shocks on trade networks allows us to capture the effect of trade via third party countries (fig.\,\ref{fig:shock}b)) or a even higher number of intermediaries.
Our model accounts for two additional shock propagation channels that result from the possibility to convert products into other products. Losses of a product, $i$, in a country, $d$, can occur if country $d$, can no longer import a product $j$, that it relies on to locally produce product $i$, (fig.\,\ref{fig:shock}c)). In addition, such losses can occur if a trading partner, $c$, lacks an input to produce a product, $i$, and reduces its exports of product $i$ to country $d$ (fig.\,\ref{fig:shock}d)).

\paragraph{Dynamic unfolding of different types of shock transmission}

\begin{figure*}[p]
    \centering
    \includegraphics[width=0.75\textwidth]{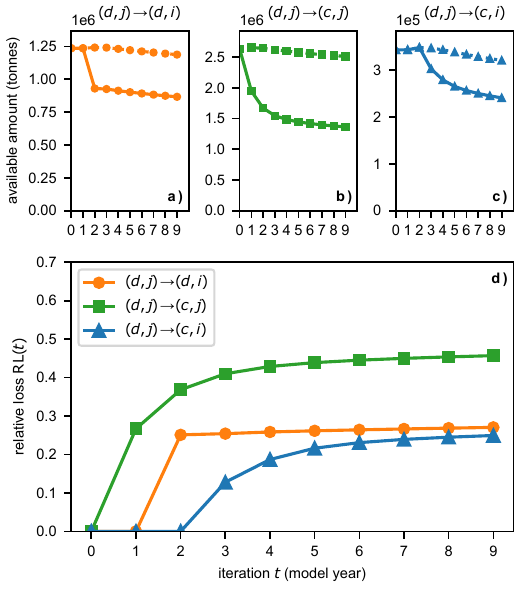}
    \vspace{0.4cm}
    \caption{Available amount (\textbf{a)}-\textbf{c)} and relative losses (\textbf{d)}) as model time passes for three different types of shock propagation channels. In each panel we consider a shock to a product, $i=\mathtt{maize}$, in a country, Ukraine $d=\mathtt{UKR}$, and the ensuing reduction of the available amount in the shocked scenario (solid line) with respect to the baseline scenario (dotted line). We show results \textbf{a)} for a different product, $i=\mathtt{poultry}$, in the same country, $d$, \textbf{b)} for the same product $j$ in another country, Portugal $c=\mathtt{PRT}$, and \textbf{c)} for a different product $i$ in a different country $c$}.
    \label{fig:dynamics}
\end{figure*}

In general, a shock $(d,j)$ to product $j$ in country $d$ can lead to three different types of impact, namely on (i) different products in the same country (local production), (ii) the same product in different countries (direct or indirect trade), or (iii) different products in different countries.
Figure~\ref{fig:dynamics} shows for the example of shocking product $j=\mathtt{maize}$ in country $d=\mathtt{UKR}$ (Ukraine), the time evolution of the available amount in a baseline case without shock, $\underline{x}_c^i(t)$, and in the shocked scenario, $x_c^i(t)$, next to the resulting relative loss, $\mathrm{RL}^{ji}_{dc}(t)$. 
First (different product, same country), this shock can reduce the amount of another product, shown here is $i=\mathtt{poultry}$, in the same country $d$ (fig.~\ref{fig:dynamics}a)). Second (same product, different country), the shock propagates through trade relations and reduces the availability of the same product, $j=\mathtt{maize}$, in another country, Portugal $c=\mathtt{PRT}$ (fig.~\ref{fig:dynamics}b)). Third and finally (different product, different country), losses of different products, shown again for $i=\mathtt{poultry}$, can occur in different countries, e.g., $c=\mathtt{PRT}$, either through reduced production in $c$ or in other countries (fig.~\ref{fig:dynamics}c)). 
Note, that the onset of losses is delayed, if the shock propagates through multiple production processes and trade (fig.~\ref{fig:dynamics}d)). In this example it may take several years until the full production losses after a shock have been realized via all direct and indirect shock transmission channels. Further details of both scenarios are described in the methods section.\\
We characterize the network topology of those layers of the trade network that are most central to our analysis in the supplementary information (SI) and show that Ukraine occupies a prominent position among the exporters. Here, we focus on shocks to the agricultural production of Ukraine but the effects of shocks to a product of choice in other countries can be explored in our interactive online data visualization~\cite{Vis-2022}.

\paragraph{The role of Ukraine in the global food system}

\begin{figure*}[h]
    \centering
    \includegraphics[width=0.9\textwidth]{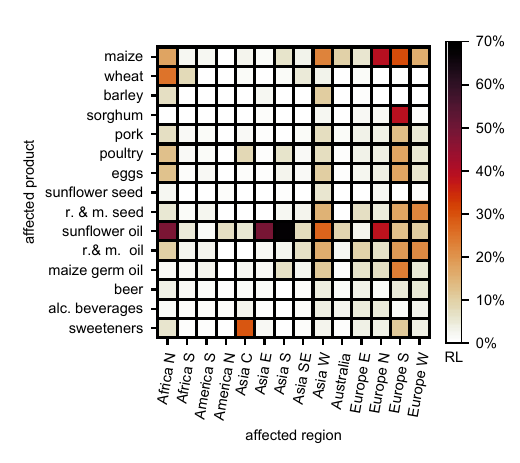}
    \caption{A simultaneous shock to the production of all food products in Ukraine affects different world regions and products. The relative loss of a given, affected product in an affected region is color-coded. The largest losses occur for sunflower oil in Southern Asia amounting to $67.8\%$ of the total production without shock. Parts of Europe, North Africa and Western Asia suffer losses in a diverse range of products including grains, edible oils, animal and luxury products, while the Americas and Australia are affected to a somewhat weaker degree. Sunflower oil stands out as the product most severely affected, but substantial losses also occur for other edible oils and grains. Rapeseed and mustardseed is abreviated r.\& m. seed, and alcoholic beverages as alc. beverages.}
    \label{fig:complete_shock}
\end{figure*}

We examine the losses that occur after a simultaneous shock to the production of all food products, $\mathcal{F}$, in Ukraine, $\mathtt{UKR}$. The resulting losses, $\mathrm{RL}^{\mathcal{F},i}_{\mathtt{UKR},C}$, of different products, $i$, in different world regions, $C$, are shown in fig.\,\ref{fig:complete_shock}.
The availability of sunflower oil is substantially reduced in several world regions. The two most strongly affected regions are located in Asia, with relative losses of $67.8\%$ arising in Southern and $48.8\%$ in Eastern Asia. Western Asia ranks fourth with relative losses of $27.1\%$. The third most affected region, Northern Africa, suffers losses of $48.3\%$. The effect on Europe is felt most intensely in the north ($38.23\%$) and less so in the south ($12.5\%$), west ($10.3\%$) and east ($2.3\%$). In the latter case we exclude the losses occurring in Ukraine itself. However, in contrast to Asian regions, Europe and Africa are also affected in their availability of other edible oils, such as rape seed and mustard seed oil (up to $21.1\%$) or maize germ oil (up to $23.0\%$).\\
The shock in Ukraine also leads to considerable losses of maize in many world regions. Northern and Southern Europe are hit strongest with losses of $39.1\%$ and $30.1\%$, respectively, followed by Western Asia with $22.2\%$ and Northern Africa $17.1\%$. The latter also faces a relative loss of $24.7\%$ of wheat.\\
Substantial losses also occur for animal products such as poultry meat. Southern Europe suffers losses of $17.2\%$ of poultry and $12.9\%$ of pork. Northern Africa loses $12.4\%$ and $6.6\%$ of the respective products. Losses reach $8.0\%$ ($1.3\%$) of poultry (pig) meat in central and $6.8\%$ ($7.0\%$) in Western Asia.\\
Regions differ considerably by the number of products for which they exhibit a direct or indirect dependence on Ukraine. Southern Europe is strongly affected, with $19$ out of $125$ products having losses of more than $10\%$, followed by Western Asia and Northern Africa, where this is the case for $15$ and $11$ products, respectively. In contrast, North America and Australia are least affected with only $5$ and $7$ out of $125$ products with a relative loss that exceeds $1\%$.\\
\begin{figure*}[h]
    \centering
    \includegraphics[width=0.9\textwidth]{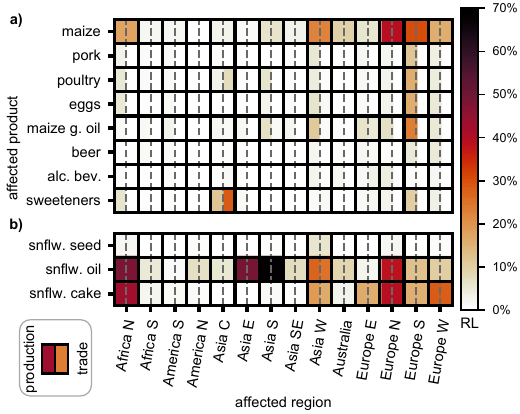}
    \caption{Comparison of losses within one layer (trade) and across layers (production). Each cell displays the loss of an affected product $i$ in a specific region after a shock in Ukraine and is split into two halves. The left half represents contributions to the losses that stem from a shock to products  other than product $i$, which can roughly be identified with contributions from production processes across different network layers. The right half shows contributions from a shock to product $i$ itself that are mostly mediated via relationships in a single trade layer. Panel (\textbf{a)}) shows the losses after a shock to Ukrainian maize production, for which we observe production-related losses in several other food products. Panel (\textbf{b)}) displays the losses after a shock to Ukrainian sunflower seeds. In this case, losses via production and trade, respectively, are of similar size. Alcoholic beverages are abbreviated as alc. beverages, sunflower as snflw., and maize germ oil as maize g. oil.}
    \label{fig:comparison}
\end{figure*}
In the following, we assess the role of production versus trade in the shock propagation. We compare the relative loss of different products, $i$, after two types of shock. On the one hand, we shock a fixed product $j$ in Ukraine and compute the relative loss, $\mathrm{RL}^{ji}_{\mathtt{URK},C}$, of other products, $i$, in different world regions, $C$. On the other hand, we shock the same products, $i$, in Ukraine and monitor the losses for $i$ in other regions, i.e. we compute $RL^{ii}_{\mathtt{UKR},C}$. The former quantifies the losses that arise on a different layer and therefore involve the conversion of products into other products, while the latter quantifies the effect within one layer, i.e. international trade. While a shocked country can still reexport products that it imported, we found that the share of reexports is low in case of Ukraine and the within layer effect constitutes a good measure for trade-related losses, see also SI.

\paragraph{Downstream impacts of a shock to Ukrainian maize production}~\\
In fig.~\ref{fig:comparison}a) these production- and trade-related contributions to the relative losses are shown for a shock to the Ukrainian maize production. Colors reflect the size of losses and each cell describes the losses of an affected product, $i$, in an affected region, $C$. Cells are split into two parts. The left half captures the production-related losses after a shock to Ukrainian maize, $\mathrm{RL}^{\mathtt{maize},i}_{\mathtt{UKR},C}$, and therefore quantifies an effect across layers. The right half, on the other hand, captures the losses after a trade-related shock in Ukraine to product $i$ itself, $\mathrm{RL}^{ii}_{\mathtt{UKR},C}$, and therefore quantifies the importance of Ukraine within one layer.\\
For the product maize itself both sides are equal by definition. The strongest effects of maize in Ukraine on maize in other countries occur in Northern Europe with losses of $39.1\%$, but Southern Europe ($30.1\%$), Western Asia ($22.2\%$) and Northern Africa ($17.1\%$) are affected as well. Note that these losses incorporate not only the lack of maize that is directly imported from Ukraine, but also reduced domestic production due to lack of seeds and the trade with third party countries, that might also rely on imports from Ukraine.\\
In addition, the upper part of fig\,\ref{fig:comparison} shows that the shock to Ukrainian maize influences the availability of pig and poultry meat in Europe, Northern Africa and Western Asia. For poultry meat the relative loss after a shock to maize in Ukraine amounts to $15.4\%$  in Southern Europe, $4.9\%$ in Northern Africa and $3.9\%$ in Western Asia. This contrasts the losses after a shock to the Ukrainian poultry meat production, which stay below $1\%$ in these world regions. This indicates that the losses in these world regions arise from a lack of fodder maize in the domestic poultry meat production and not from a trade of poultry meat with Ukraine. A similar pattern can be observed in the supply of other products that rely on maize as an input to production.\\
The availability of sweeteners in central Asia is an exception. Here the effect within one layer, with losses of $28.3\%$, is larger than the effect across layers, with a loss of $11.1\%$, after a shock to the Ukrainian maize production. Similarly, the relative loss of alcoholic beverages in central Asia after a shock to this product in Ukraine amounts to $1.5\%$ and exceed the losses of $0.7\%$, that occur after a shock of the Ukrainian maize production. This situation arises as both products can be made from a variety of input products other than maize.
\paragraph{Ukraine is a critical supplier of sunflower oil}~\\
Fig.~\ref{fig:comparison}b) shows results for a shock to the sunflower seed production of Ukraine.
In contrast to the maize shock in fig.~\ref{fig:comparison}a) the losses induced across layers (left half) and within the same layer (right half) are of similar size.
In Southern and Eastern Asia relative losses of $67.7\%$ and $48.7\%$ of sunflower oil are observed for a shock to the Ukrainian sunflower seed production. A shock to Ukrainian sunflower oil leads to equal losses in these world regions. Losses of similar size occur in Northern Africa, which loses $48.2\%$ due to a shock to sunflower seeds and $48.1\%$ after a shock to sunflower oil, and Northern Europe with losses of $38.2\%$ and  $38.1\%$ respectively. The difference between the two types of shock is largest in Western Asia where losses of $27.0\%$ of sunflower oil arise after a shock to Ukrainian sunflower seeds and $24.8\%$ after a shock to Ukrainian sunflower oil. As a shock to Ukrainian sunflower seeds also causes losses of $5.2\%$ of sunflower seeds in this region, the additional losses are likely due to a reduced local production of oil as a result of the reduced availability of sunflower seeds.
Most world regions also suffer losses of sunflower cake, a residue from oil seed crushing that can be used as fodder. The largest losses occur in Western Africa $42.9\%$, Northern Europe $39.1\%$ and Western Europe $28.1\%$ for a shock to Ukrainian sunflower seeds. For a shock to Ukrainian sunflower cake itself losses are slightly lower or equal, amounting to $42.4\%$, $39.1\%$ and $28.1\%$ in the respective world regions.
The relative losses of sunflower seeds in all world regions are much weaker than the losses of sunflower oil.
Our multilayer modelling framework can readily be applied to other types of shocks, which we demonstrate in the SI within a case study to identify dependencies of the German pork production.

\FloatBarrier
\section{Discussion}\label{s:discussion}

We propose a multilayer network model that accounts for production dependencies among $125$ food and agricultural products, allowing us to assess the losses of food products around the globe after a localized shock. The model takes into account trade dependencies as well as the conversion of products into other products along the global supply chain. We use data on production and trade of food and agricultural products to determine the parameters of the model. We introduce the relative loss, $\mathrm{RL}$, that describes the multidimensional dependencies of the availability of various food products in each country with respect to shocks to food products in any other country in an easily interpretable way.~\\~\\
First, we show that a complete shock to the Ukrainian production leads to stark losses to grains and dietary oils in parts of Europe, Western Asia and Northern Africa. It seems plausible to assume that high income countries will be able to mitigate the losses by tapping into reserves or switching suppliers in spite of possible higher costs, while middle and low income countries are likely unable to mitigate the losses.
In Northern Africa and Western Asia food security was already critical prior to the onset of the war due to factors, that vary on a country by country basis but include the aftermath of supply-chain disruption due to the COVID-19 pandemic, extreme weather events and local conflict. Our results corroborate the increased risk of a food crisis in these regions~\cite{Abay_2023-03,Behnassi_2022-06}. 
Large losses in grain supply are likely to have direct adverse affects on food security given the large shares of grains in the diets of countries in these regions, e.g., wheat makes up about $35.3\,\%$ of the caloric intake in Egypt and $39.6\,\%$ in Syria~\cite{Mottaleb_2022-12} and is also an important protein source~\cite{Shiferaw_2013-06}.
Even in countries in which food security is not directly threatened, supply disruptions can make healthy and nutritious food less affordable, thereby forcing low income households to switch to less nutritious food options~\cite{Behnassi_2022-06}.
Overall, these findings highlight the importance of Ukraine as a critical supplier of agricultural products and serve as an example for the fragility of global supply chains with respect to the failure of key producers due to geopolitical instability.\\~\\
Second, we investigate the effects of shocks to single products in Ukraine on several other products in different world regions. We compare the effects of a shock to an important input product on more processed products, i.e., shock propagation across layers, to the effects of shocks to the processed products themselves purple, i.e., shock propagation within the same layer. We find that the production of pork and poultry in Southern Europe, Western Asia and Northern Africa relies heavily on the imports of fodder maize from Ukraine but that the role of direct meat imports from Ukraine is small. This means the shock propagation channel shown in fig.~\ref{fig:shock}c), where countries import a shock along with inputs they require for their domestic food production, contributes considerable to the losses.\\
In contrast, shocks to sunflower seeds, the input product, and to sunflower oil, the processed product, lead to similar losses abroad. Shock propagation is likely dominated by the channel displayed in fig.~\ref{fig:shock}d), i.e. losses in inputs to domestic production processes that propagate to other countries via exports of the processed product. As we measure relative and not absolute loss, we can identify a substantial reduction of the availability of sunflower oil in South and East Asia. While these are not the largest importers of Ukrainian sunflower oil in absolute terms, a large share of their sunflower oil imports stems from Ukraine. This shows how non-diversified supply relations can lead to large relative losses in case of a shock.
Both cases highlight the importance of integrated trade and production processes in the assessment of losses. Shocks to input products in a given country can propagate to other product layers, either through the production processes in the same country or after trade in other countries, thereby causing losses in downstream products. This is especially apparent in the case of maize, as the crop does not only substantially contribute to the daily diet in many countries, both in terms of caloric intake and protein supply, but also serves many other use cases. The latter include on the one hand feed for life stock and on the other hand the production of bioenergy~\cite{Shiferaw_2011-09}.
Balancing these conflicting use-cases and associated (economic) interests poses a challenge for the relevant stakeholders. Our analysis raises attention for this type of interdependence and calls to address the associated challenges.
From a methodological perspective, we emphasize that losses downstream in the supply-chain can be obscured in a risk analysis based on trade data alone. This underscores the need to routinely extend risk analysis along more than one product dimension in global production networks.\\~\\
These examples illustrate the wide-ranging consequences of the ongoing war in Ukraine and how our model can be used to assess supply disruptions that can reduce the availability of nutritious food. Additional mechanisms, such as price increases, can shift the burden further upon vulnerable populations, thereby exacerbating existing food insecurity and nutritional challenges~\cite{Osendarp_2022-04}. Our study shows how a spatially confined shock does not only affect the local population, but can lead to global supply disruptions. It also shows that these adverse effects extend along the entire supply chain, from the cultivation of crops to the production of food and animal feed, to the life stock sector, all the way to the consumer.
Managing the risk along the food supply chain resulting from the conflict in Ukraine requires a multi-faceted perspective, that takes both direct and indirect effects into account. Our results enhance supply chain visibility and transparency. This can help identifying potential risks and disruptions in the food system and developing contingency plans that enable food supply chain stakeholders prepare for potential disruptions and mitigate their impact. For countries that heavily depend on Ukrainian imports ---either directly or indirectly--- this could for example mean diversifying their supply of imports. Viewing the food supply chain as network with interdependent layers can inform policies that rely on local interventions and resulting ripple effects to counter existing crises. For example, it suggests that positive effects of providing seed material and assuring access to means of production can extend beyond the directly supported crop producers. It also suggests that care has to be taken when allocating resources to different use cases and that it might be necessary to adapt allocation strategies during a crisis, e.g. by using certain crops directly as food rather than as input for further production processes. In all cases, countries should strive to minimize food waste. Reexamining previous work~\cite{Burkholz_2019-10,Gomez_2020-09} showing that cascading export restrictions can exacerbate shortages in the light of inter-product dependencies makes clear that export bans on single products could potentially lead to unintended side effects on other products, worsening the situation even further.
\\~\\
While existing studies of the international food system have created valuable insights, our work addresses various open challenges in the field:
Previous shock propagation frameworks were limited to a few products~\cite{DAmour_2016-02, Burkholz_2019-10, Heslin_2020, Naqvi_2021-10} or aggregated them in terms of calories~\cite{Marchand_2016-09, Grassia_2022-03}, monetary units~\cite{Gomez_2020-09} or virtual water~\cite{Sartori_2015-11,Tamea_2016-01}. Here we extended a shock propagation to $125$ food and agricultural products.
Even more importantly, we took into account the production dependencies among these products, while previous work considered products in isolation. This includes studies that simulate the international propagation of food supply shocks~\cite{DAmour_2016-02,Burkholz_2019-10,Marchand_2016-09,Heslin_2020, Grassia_2022-03}, analyze historical shocks to production~\cite{Distefano_2018-08, Gephart_2017-01} as well as descriptions~\cite{BerfinKarakoc_2021} and predictions~\cite{Fair_2017-08} of the food trade network topology.
Finally, the explicit representation of production processes opens up the possibility to customize production functions if sufficient data on the production process of individual products is available. This sets our work apart from existing shock propagation frameworks that work directly with input-output tables~\cite{Grassia_2022-03}.~\\~\\
Nonetheless, our study faces several limitations:
The proposed model provides a simple, mechanistic picture of the dynamics of food trade and production. We focus on the conversion of products into each other and do not consider impacts of stocks and changing resource allocation based on availability. Furthermore, we simulate dynamics that take place on a static multi-layer network. As a result, we do not capture the dynamical restructuring of supply relations that are likely to occur after a shock. This concerns on the one hand trade relations, in which a failing exporter may be replaced by a competitor on the market, and on the other hand production relations, in which production recipes can be adapted to some extent based on the available products. These issues are tightly connected to prices and markets, which are not included in our mechanistic model.
The above effects have been shown to redistribute the burden of the losses~\cite{Marchand_2016-09,Gephart_2016-02}, as countries that have stocks available or are wealthy enough to buy resources at higher prices can buffer shocks, while countries that lack these capabilities might be affected more strongly. Our model represents a worst case scenario of a complete and sustained production loss and is likely to overestimate real losses. Nevertheless, our results might inform crisis response by providing estimates for the losses incurred in individual countries unless they make substitution efforts or consider other mitigation measures.\\
We restrict ourselves to the case of linear production functions. The incorporation and comparison of other production functions opens up possibilities for future research. Work on shocks in firm-level supply chain networks suggests that non-linear production functions of the Leontief type can lead to larger losses compared to situations with linear production functions~\cite{Diem_2022-05}. We expect this effect to influence our results particularly in cases where a production recipe includes a comparably small but essential amount of an input, e.g. hops for beer production.\\
Furthermore, the data has limits in terms of spatial resolution as well as granularity and scope of products. On the one hand, data on the subnational level would allow us a more realistic assessment of losses. This is especially true for shocks like natural disasters or extreme weather events that are spatially localized but do not stop at national borders. On the other hand, finer resolution on the product level would enable us to better trace shocks through the process of production. The product wheat, for example, encompasses also certain wheat products like bread and pasta. Extending the model to non-food input products, such as fertilizers~\cite{Obersteiner_2013-11,Barbieri_2022-02} or machinery, is necessary to create a more complete picture of a country's dependencies.\\
We have chosen to base our model on shocks to products to reflect losses that occur if the product itself or necessary means of production are destroyed, e.g. by armed conflict or natural disasters. However, other scenarios such as blockage of exports either because of an export ban~\cite{Burkholz_2019-10} or physical blockade of infrastructure, are conceivable and could be incorporated into the model. In context of the war in Ukraine, both scenarios contribute. On the one hand a naval blockade makes it impossible to export maize and wheat, on the other hand the agricultural production itself is on halt as farmers serve as soldiers, fields are mined and infrastructure is destroyed~\cite{Bubola_2022-04,Reed_2022-04}.\\~\\
While food security extends far beyond the trade and production centered perspective adopted herein and incorporates social, cultural and ecological aspects~\cite{Reichstein_2021-04,Gaupp_2020-06, DeRaymond_2021-06}, production and trade are essential components to secure the basic availability of food. While we focused here on the ongoing war in Ukraine, armed conflict is only one among many possible causes of shocks to the global food system, as in an age of climate change, crop failures due to droughts, other extreme weather events or food pests become increasingly likely. Tools from complexity and network science, as those presented herein, enable us to uncover direct and indirect dependencies along the global supply chain~\cite{Davis_2021-01}, in these and other scenarios. Raising awareness for the multidimensional systemic risk incorporated in the global trade and production of food, can help to design more resilient supply chains.

\section{Methods}\label{s:methods}
\paragraph{Food production and trade as an iterative three-step process}~\\
The model consists of $192$ countries and territories that trade and produce $125$ food and agricultural products. The set of countries is denoted $\mathcal{C}$ and the set of products $\mathcal{F}$. Production is modelled by $118$ types of production processes, which form a set, $\mathcal{P}$. This corresponds to the countries and territories, products and production processes of the food and agriculture biomass input--output (FABIO) model~\cite{Bruckner_2019-10}. The mapping of countries and territories to world regions can be found in the supplementary information. A python implementation of the model together with input data in a suitable format is publicly available.~\cite{github-2023}
A country, $c$, possesses a total amount, $x_c^i(t)$, of a product, $i$, at iteration $t$. This amount is split in fixed, but country- and product-specific fractions, $\eta_{c,i}^{k}$, to serve different purposes, $k$ (here and in the following, index variables used as superscripts denote indexes, not exponents). These purposes are exports, $\eta_{c,i}^\mathrm{exp}$, input to production, $\eta_{c,i}^\mathrm{prod}$, consumption as food, $\eta_{c,i}^\mathrm{food}$, and other uses, $\eta_{c,i}^\mathrm{else}$, including waste, non-food consumption and stockpiling. The amount, $x_c^i(t)$, changes as countries repeatedly undergo the three-step process of production, trade and allocation to different purposes.\\
In the first step, countries rely on different types of production processes to create and to convert products into each other. For each type of process, $k$, in each country, $c$, there exists a function, $f_{c,i}^{k}(p_c(t-1))$, that maps the available amounts of inputs, $p_c(t-1)$, from the previous time step to the produced amount of output product, $i$, supplied by this process. Here $p_c(t-1)$ denotes a vector with components $p_c^j(t-1) = \eta_{c,j}^\mathrm{prod} x_c^j(t-1)$. As different processes can supply the same output product, the total amount $o_c^i(t)$ of product $i$, produced in country $c$ at iteration $t$ is

\begin{equation}
    o_c^i(t) = \sum_{k\in\mathcal{P}} f^{k}_{c,i}(p_c(t-1))~.
\end{equation}

An example of a product that is supplied by multiple processes are animal fats or animal hides, which accrue in the slaughter of different types of animals. \\ 
Moreover, it is possible that different types of processes use the same input product. Therefore, it is necessary to split the amount, $p_c^j(t-1)$, of product $j$, that is available for production in a country, $c$, among the processes that use it. The fraction assigned to a particular process, $k$, is denoted $\nu_{c,j}^k$ and depends on the country, $c$, the product, $i$, and the process type, $k$. This case occurs for example in the farming of poultry and pigs, which can both be fed with maize. The assumption that the input products are completely substitutable, as long as they are from the set of valid input products, $\mathcal{I}_{c,k}$, to the production process, $k$, in country, $c$, motivates 

\begin{equation}
    f^k_{c,i}(p_c(t-1)) = \alpha_{c,i}^k \sum_{j \in \mathcal{I}_{c,k}} \nu_{c,j}^k p^j_c(t-1) + \beta_{c,i}^k~,
\end{equation}

as a mathematical form for the amount of a product, $i$, produced by a process, $k$, in a country, $c$. Here, $\alpha_{c,i}^k$ is the amount of output product, $i$, that can be produced by process $k$ in country $c$ for a given amount of input and $\beta_{c,i}^k$ is the amount of product $i$ produced in absence of any input. In the present initiation of the model either $\beta_{c,i}^k$ or $\alpha_{c,i}^k$ is non-zero, but not both. The production of apples is an example for a process with non-zero $\beta_{c,i}^k$. In other cases non-zero values of $\beta_{c,i}^k$ occur because countries report outputs of a process even though it is reasonable to assume that the process would rely on inputs.\\ 
In a second step of iteration, $t$, each country, $d$, trades the amount $e_d^i(t-1)$ of product $i$ that it reserved for this purpose in the previous iteration, $t-1$. The trade of each product, $i$, is governed by a weighted, directed network described by a matrix, $T^i$, and countries as nodes. An entry, $T^i_{cd}$, of this matrix indicates the share of country $d$'s exports that country $c$ receives. This allows us to write the imports, $h^i_c(t)$, of product $i$ of country $c$ in time step $t$ as

\begin{equation}
    h^i_c(t) = \sum_{d\in\mathcal{C}} T_{cd}^i e_d^i(t-1).
\end{equation}

The trade relations between countries and the production dependencies among products can thus be represented as a bipartite multi-layer network. The trade network for each product forms a single layer with countries as nodes and the links indicate trade relations. The replica nodes, representing the same country on different layers, are connected by directed interlayer links via nodes of a second type, which represent different production processes. A country node in a layer representing the trade of a product is in-neighbor of a specific process in that country if this product is an input to the production process and it is an out-neighbor if the product is an output of the production process.\\
In the third step, allocation, each country, $c$, assigns fractions of the total available amount of product $i$, $x_c^i(t)=o_c^i(t)+h_c^i(t)$, obtained either from domestic production or imports from other countries to different purposes according to

\begin{align}
    p_c^i(t) &= \eta_{c,i}^\mathrm{prod}x_c^i(t) \\
    e_c^i(t) &= \eta_{c,i}^\mathrm{exp}x_c^i(t) \\
    k_c^i(t) &= \eta_{c,i}^\mathrm{food}x_c^i(t) \\
    r_c^i(t) &= \eta_{c,i}^\mathrm{else}x_c^i(t)~,
\end{align}

where $k_c^i(t)$ denotes the amount available of product $i$ in country $c$ for consumption as food and $r_c^i(t)$ is the amount of product $i$ in country $c$ that is allocated to all other use cases. Thereafter the accounting variable, $x^i_c(t)$, is reset and the next time step begins with the first sub-step, production.
We calibrate the parameters and initial conditions, i.e. available amount at time-step $t=0$ of our model to trade and production data of the year $\tau=2013$, as described below. Note, that while the topology of the trade network, the fractions of goods assigned to different purposes and the production functions, stay fixed during simulation, the total available amount of each product ---and as a consequence also the amount assigned to each purpose--- changes with each iteration $t$. This gives rise to dynamics on a static network. An overview over parameters and variables, how they are obtained and whether they are static or time-dependent can be found in the SI.

\paragraph{The relative loss captures the effects of a shock}~\\
To determine the effects that a shock to the production of certain products in certain countries has on the availability of all other products in all other countries, we compare a baseline scenario to a shocked scenario. The baseline scenario consists of iterating the dynamics described above for $t_\mathrm{end}=10$ time steps. In the shocked scenario a fraction $\phi$ of the production output, $o_d^j(t)$, of the shocked product, $j$, in the shocked country, $d$, is destroyed after the production step. This means only an amount

\begin{equation}\label{eq:prod_shock}
    (1-\phi)o_d^j(t)
\end{equation}

is available after production at all times, $t\in\{0,\dots, t_\mathrm{end} \}$. Here we investigate a worst case scenario, in which the complete production is lost, i.e., $\phi=1$, and this production failure persists over all iterations. A shock to more than one product is realized by applying eq.~\ref{eq:prod_shock} to all products, $j$, from a set of shocked products. The same is true for shocks to more than one country.\\
The relative loss, $\mathrm{RL}^{ji}_{dc}$, of product $i$ in country $c$ in response to a shock to the production of product $j$ in country $d$ is a measure for the severity of the adverse effects on a country and defined as the relative reduction in total available amount in the shocked with respect to the baseline scenario,

\begin{equation}\label{eq:RL}
    \mathrm{RL}^{ji}_{dc}(t) = \frac{\underline{x}^i_c(t) -X^{ji}_{dc}(t)}{\underline{x}^i_c(t)}~, 
\end{equation}

with $X^{ji}_{dc}(t)$ denoting the total amount of product $i$ available in a country, $c$, at time step $t$ after a shock to product $j$ in country $d$ and $\underline{x}^i_c(t)$ denoting the total amount of product $i$ available in country $c$ at time $t$ in the baseline scenario. We drop the time dependence when referring to the relative loss in the last time step $\mathrm{RL}^{ji}_{dc} = \mathrm{RL}^{ji}_{dc}(\mathrm{t_\mathrm{end}})$. To assess the effect of a shock on a whole region, $C$, that consists of several countries, we aggregate the available amount of a product, $i$, on the regional level

\begin{equation}
    \mathrm{RL}^{ji}_{dC} = \frac{\sum_{c\in C} \underline{x}_c^i({t_\mathrm{end}}) - \sum_{c\in C} X^{ji}_{dc}({t_\mathrm{end}})}{\sum_{c\in C} \underline{x}_c^i({t_\mathrm{end}})}~.
\end{equation}
\paragraph{Supply and use tables of the FABIO model}~\\
The free parameters of the model are determined from the supply and use tables of the FABIO model~\cite{Bruckner_2019-10} for the year 2013. The FABIO model is a multi-regional input--output model in physical units. The entry $S_{(c,k),(d,i)}$ of the supply table, $S$, describes the amount of product $i$ a production process, $k$, supplies in country $d$. Processes only supply to the country in which they are based and therefore the supply table is country-block-diagonal, i.e.

\begin{equation}
   c\neq d \Rightarrow S_{(c,k),(d,i)}=0 ~~~\forall\,i \in\mathcal{F}\wedge k\in\mathcal{P}~. 
\end{equation}

The entry $U_{(d,i),(c,k)}$ of the use table, $U$, gives the amount of product $i$ originating from country $d$ that process $k$ in country $c$ uses. Countries can use products from different countries and therefore the use table, $U$, has no special block structure.\\
The demand table entry $Y_{(d,i),(c,l)}$ specifies the amount of product $i$ from country $d$ that country $c$ uses to satisfy the demand for the purpose $l$. These purposes are labeled food, losses, stock addition, other, unspecified and balancing. They are summarized in the set $\mathcal{V}$. We denote the positive part of the demand, $Y$, as $Y^+$ and the negative part as $Y^-$. The latter can only be nonzero if $l\in\{\text{balancing},\text{stock addition}\}$. The sets of products, countries and processes are adopted directly from the FABIO model, as described in~\cite{Bruckner_2019-10}.
\paragraph{Trade networks}~\\
The entry $T^i_{cd}$ of the matrix $T^i$ describes the share of exports of product $i$ from country $d$ that are directed towards country $c$. It is determined as the amount of a product, $i$, originating from country $d$ used either by any production process, $k$, in country $c$ or to satisfy any demand-purpose, $l$, in country $c$ relative to the amount of product $i$ from country $d$ that is used by any production process, $k$, or any demand purpose, $l$, in any country other than $d$ itself, i.e. 

\begin{equation}
    T^i_{cd}= \frac{\sum_{k\in\mathcal{P}} U_{(d,i),(c,k)}  + \sum_{l\in\mathcal{V}}Y^+_{(d,i),(c,l)}}{\sum_{k\in\mathcal{P},c\neq d} U_{(d,i),(c,k)}  + \sum_{c\neq d,l\in\mathcal{V}}Y^+_{(d,i),(c,l)}}
\end{equation}

for $c\neq d$ and $T^i_{cd}=0$ if $c=d$. By definition the entries satisfy $T^i_{cd}\in[0,1]$ and the out-strength of a country is always

\begin{equation}
    \sum_{c\neq d} T^{i}_{cd} = 1 ~~\forall~d \in \mathcal{C} \wedge i\in\mathcal{F}.
\end{equation}

\paragraph{Allocation to use cases}~\\
To avoid negative demands, $Y^-$, we introduce a balancing term for each country, $c$, and product, $i$, as the difference between the supply of the product, $i$, by any local process and the sum of the domestic or foreign use of the product by any process and the positive part of the domestic or foreign demand for the product for any purpose, 

\begin{equation}
    B_{(c,i)} = \sum_{p\in\mathcal{P}}S_{(c,p),(c,i)} - \sum_{d\in\mathcal{C},p\in\mathcal{P}} U_{(c,i),(d,p)} - \sum_{d\in\mathcal{C},k\in\mathcal{V}}Y^+_{(c,i),(d,k)}~.
\end{equation}

The positive part of the balancing, $B$, is denoted $B^+$ and its negative part as $B^-$. With this balanced demand it is possible to determine the fractions, $\eta_{c,i}^k$, that are used in the allocation step. The share of products used as input to production, $\eta^\mathrm{prod}_{c,i}$, is the amount of product $i$ originating from any foreign country and used in any local process in country $c$ divided by the sum of this amount and the amount of product $i$ originating from any country and used to satisfy any type of demand in country $c$ and the balancing term for product $i$ in country $c$, 

\begin{equation}
    \eta_{c,i}^\mathrm{prod} = \frac{ \sum_{k\in\mathcal{P},d\in\mathcal{C}}U_{(d,i),(c,k)}}{\sum_{k\in\mathcal{P},d\in\mathcal{C}}U_{(d,i),(c,k)} + \sum_{d\in\mathcal{C}, i \in \mathcal{F},l\in\mathcal{V}}Y^+_{(d,i),(c,l)}+B^+_{(c,i)}}~.
\end{equation}

The fraction, $\eta^\mathrm{food}_{c,i}$, of product $i$ used to satisfy the demand for food in a country, $c$, is obtained by dividing the amount of product $i$ originating in any country and used to satisfy food demand in country $c$ by the same denominator,

\begin{equation}
    \eta_{c,i}^\mathrm{food} = \frac{\sum_{d\in\mathcal{C}}Y^+_{(d,i),(c,\mathrm{food})}}{\sum_{k\in\mathcal{P},d\in\mathcal{C}} U_{(c,i),(d,k)} + \sum_{d\in\mathcal{C},l\in\mathcal{V}}Y^+_{(c,i),(d,l)}+B^+_{(c,i)}}~.
\end{equation}

The fraction, $\eta^\mathrm{else}_{c,i}$, that is used for any non-food purpose is obtained by summing the demand in the numerator over all non-food demand purposes including the balancing, i.e.

\begin{equation}
    \eta_{c,i}^\mathrm{else} = \frac{\sum_{d\in\mathcal{C},l\in\mathcal{V}\setminus \{\mathrm{food}\}}Y^+_{(d,i),(c,l)} + B^+_{c,i}}{\sum_{k\in\mathcal{P},d\in\mathcal{C}} U_{(c,i),(d,k)} + \sum_{d\in\mathcal{C},l\in\mathcal{V}}Y^+_{(c,i),(d,l)}+B^+_{(c,i)}}~.
\end{equation}

Finally the share of product $i$, country $c$ reserves for exports in the allocation step is calculated by dividing the amount of a product, $i$, originating from a country, $c$, that is used either by any production process in any foreign country or to satisfy any type of demand in any foreign country, 

\begin{equation}
    \eta_{c,i}^\mathrm{exp} = \frac{\sum_{k\in\mathcal{P},d \neq c} U_{(c,i),(d,k)}  + \sum_{d\neq c,l\in\mathcal{V}}Y^+_{(c,i),(d,l)}}{\sum_{k\in\mathcal{P},d\in\mathcal{C}} U_{(c,i),(d,k)} + \sum_{d\in\mathcal{C},l\in\mathcal{V}}Y^+_{(c,i),(d,l)}+B^+_{(c,i)}}~.
\end{equation}

By definition these fractions cover all possible use cases and thus
\begin{equation}
    \eta_{c,i}^\mathrm{prod} + \eta_{c,i}^\mathrm{food} + \eta_{c,i}^\mathrm{exp} + \eta_{c,i}^\mathrm{else} = 1~~\forall c\in\mathcal{C} \wedge i\in\mathcal{F}~.
\end{equation}

\paragraph{Properties of production processes}~\\
During the production step, the amount of a product, $i$, available for production in a country, $c$, needs to be split among the different processes that use it as an input. The corresponding fraction, $\nu^k_{c,i}$, of this amount that a process, $k$, obtains is calculated as the quotient of the amount of product $i$ originating from any country used by process $k$ in country $c$ by the amount of product $i$ originating from any country used by any process in country $c$,

\begin{equation}
    \nu_{c,i}^k = \frac{\sum_{d\in\mathcal{C}} U_{(d,i),(c,k)}}{ \sum_{d\in\mathcal{C},k\in\mathcal{P}}U_{(d,i),(c,k)}}~.
\end{equation}

As a result of this definition, no share of input products is left unused and the fractions satisfy

\begin{equation}
    \sum_{k\in\mathcal{P}} \nu_{c,i}^k = 1~~~\forall~c\in\mathcal{C} \wedge i\in\mathcal{F}~.
\end{equation}
The production process of type $k$ based in a country, $c$, is defined by coefficients, $\alpha^k_{c,i}$, that describe how much of output product, $i$, is produced per amount of available input. For each process type, $k$, in country $c$ the set of output products, $\mathcal{O}_{c,k}$, comprises those products that the process supplies,

\begin{equation}
     \mathcal{O}_{c,k} = \{i \vert S_{(c,k),(c,i)}>0~\mathrm{for}~k\in\mathcal{P}\wedge i\in\mathcal{F}\}~.
\end{equation}

The set of input products, $\mathcal{I}_{c,k}$, to process type $k$ in country $c$ consists of those products, of which a non-zero amount is used by the process

\begin{equation}
     \mathcal{I}_{c,k} = \{i~\vert~U_{(d,i),(c,k)}>0~\mathrm{for}~d\in\mathcal{C}\wedge j\in\mathcal{F}\}~.
\end{equation}

The same process type can therefore supply different products and rely on different inputs depending on the country. For example, the composition of the fodder, that animals receive in different countries, might vary.
If the process relies on inputs, $\mathcal{I}_{c,k}\neq\emptyset$, then $\beta_{c,i}^k = 0$ and 

\begin{equation}
    \alpha^k_{c,i} = \frac{S_{(c,k)(c,i)}}{\sum_{d\in\mathcal{C},j\in \mathcal{I}_{c,k}} U_{(d,j),(c,k)}}~~~\text{if}~i\in\mathcal{O}_{c,k}~,
\end{equation}

but $\alpha^k_{c,i} = 0$ if $i\notin \mathcal{O}_{c,k}$.
If the process does not rely on inputs, $\mathcal{I}_{c,k}=\emptyset$, then $\alpha_{c,i}^k=0$ and $\beta_{c,i}^k=S_{(c,k),(c,i)}$ for $i\in \mathcal{O}_{c,k}$ and $\beta_{c,i}^k=0$ for $i\notin \mathcal{O}_{c,k}$.\\
At time $t=0$ each country, $c$, starts with a total available amount, $x_c^i(0)$, of product $i$ that equals the positive domestic and foreign demand and the amount used by any foreign or domestic production process, 

\begin{equation}
    x_c^i(0) = \sum_{d\in\mathcal{C},k\in\mathcal{P}}U_{(c,i),(d,k)}+\sum_{d\in\mathcal{C},l\in\mathcal{V}}Y^+_{(c,i),(d,l)}
\end{equation}

This amount determines the initial values for production input, $p_c^i(0)$, and exports, $e_c^i(0)$. Negative demand and balancing terms are taken into account by adding the amount

\begin{equation}
    \tilde{x}_c^i = -\sum_{d\in\mathcal{C}}Y^-_{(d,i),(c,\mathrm{stock\_addition})} - B^-_{(c,i) }~.
\end{equation}

to the output $o_c^i(0)$ in the first iteration $t=0$.

\section*{Data Availability}
The data used in this study as input for simulations is available on GitHub,\\ \url{https://github.com/L-MoNi/shock-propagation-food-supply}.

\section*{Code Availability}
Python was used to perform the simulations and data analysis. Simulation and analysis code for this study is available in a repository at \url{https://github.com/L-MoNi/shock-propagation-food-supply}.

\section*{Acknowledgements}
We are grateful to Christian Diem, Tobias Reisch, Anton Pichler, Jan Hurt and Frank Neffke for illuminating discussions.
This work was supported by the Austrian Science Promotion Agency FFG under 882184 (ST)and 886360 (ST) as well as by the Austrian Science Fund FWF under P 31598-G31 (MB).

\section*{Author Contribution Statement}

Peter Klimek, Moritz Laber and Stefan Thurner contributed to study conception and design. Data was collected by Martin Bruckner. Analysis was carried out by Moritz Laber. Results were interpreted by Peter Klimek, Moritz Laber and Stefan Thurner. The artwork was conceived by Liuhuaying Yang and Moritz Laber. The first draft of the manuscript was written by Moritz Laber and all authors commented on previous versions of the manuscript. All authors read and approved the final manuscript.

\section*{Competing Interest Statement}
The authors declare no competing interests.

\appendix

\section{Pork availability in Germany depends on fodder imports}
\begin{figure*}[h]
    \centering
    \includegraphics[width=\textwidth]{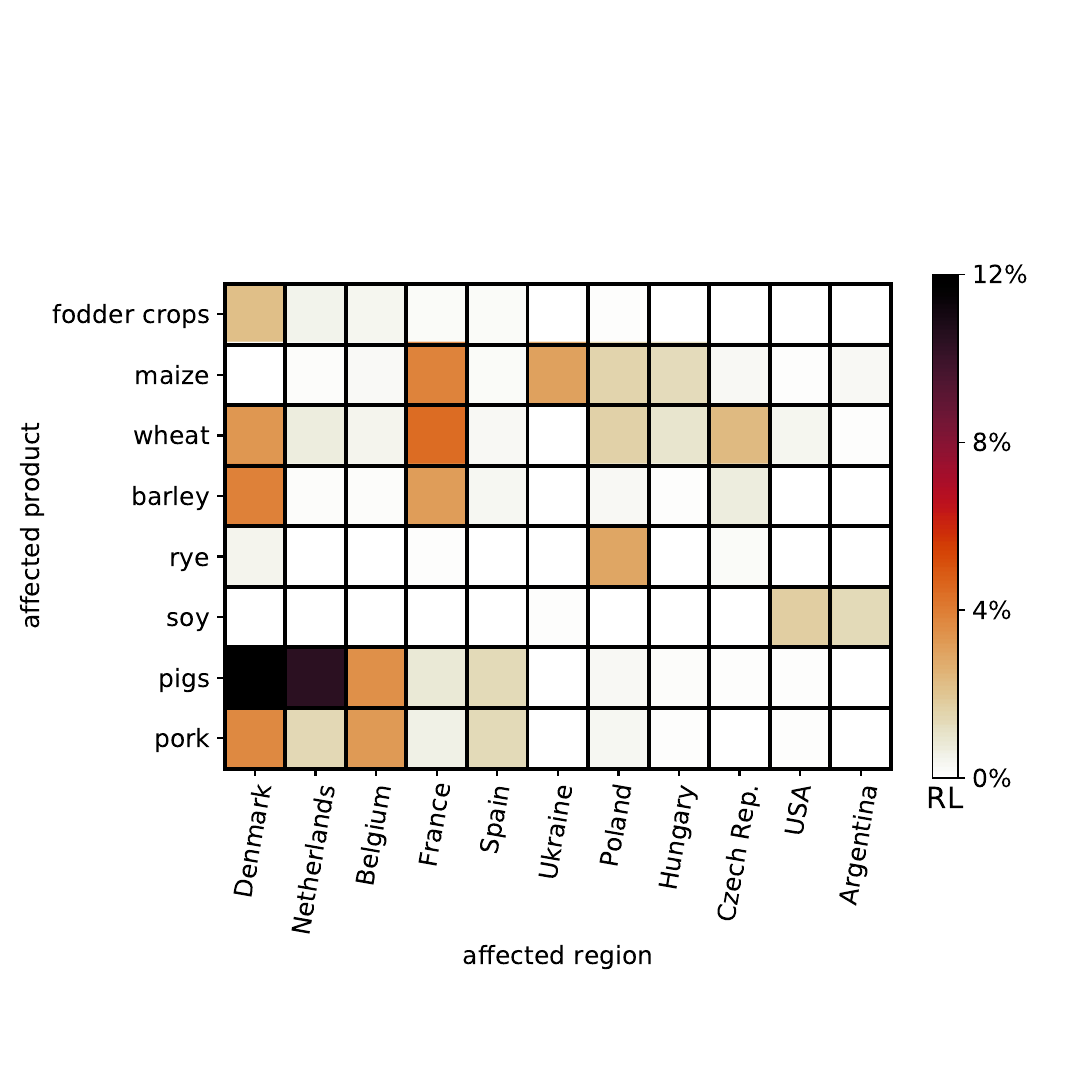}
    \caption{Losses of pork in Germany (color coded) after a shock to different products in different countries. For example, a shock to the production of pigs in Denmark induces a relative loss of $11.8\%$ in Germany. In all countries shocks to products other than pork cause larger losses then the shock to pork production. The largest losses occur for shocks to the production of life pigs in Denmark and the Netherlands. Denmark, France and several countries in Eastern Europe cause losses of pork in Germany if their grain production fails. The USA and Argentina affect the availability of German pork via their soy production.}
    \label{fig:loss_pigmeat}
\end{figure*}

~\\
To show the versatility and broad applicability of our method, we investigate the critical suppliers for the availability of pork in Germany. These are countries $c$, in which the a shock to a certain product $i$ reduces the availability of pork in Germany. Figure~\ref{fig:loss_pigmeat} displays the relative losses of pork in Germany for shocks to different products in different countries. Note, that in contrast to the main text, the country and product which receive the shock vary and the country and product, for which the relative loss is monitored remains fixed.
We find that grains, like maize, wheat, barley, and rye, from European countries is important for the availability of pork in Germany. France has the largest impact and can induce losses of $4.4\%$ via wheat, $3.8\%$ via barley and $3.1\%$ via maize production. In contrast, shocks to pig husbandry and pork production in France have influences of below $1\%$ on the availability of pork in Germany. A similar pattern can be observed for crops from Eastern European countries. We observe losses of $2\%$ to $3\%$ for a shocks to maize in Ukraine, rye in Poland and wheat in Czech Republic. The United States and Argentina influence the availability of pork via their soy exports and a shock to the soybean production in these countries causes losses of $1.7\%$ and $1.3\%$ respectively. 
Denmark and the Netherlands affect the availability mostly via different products. A shock to their domestic production of pigs causes losses of pork in Germany of $11.8\%$ and $10.4\%$ respectively. A shock to their pork production only leads to much smaller losses of
$3.7\%$ and $1.3\%$. This suggests that these countries export living pigs that are slaughtered abroad. In case of Spain and Belgium the losses Germany faces after a shocks to pigs or pork are of similar size, indicating that those countries rather export meat than live animals. 

In summary, these findings show that different countries affect the amount of pork in Germany via different input products to pork production. Important inputs include grains from France and Eastern Europe as well as soy from the United States and Argentina. The largest losses occur however for shocks to pigs in Denmark and the Netherlands. These dependencies of Germany due to missing inputs to production provide an example for shock propagation along the channel shown in fig.~2c) in the main text. As the influence of each supplier varies with the shocked product, the full information about the risk of the supply relation cannot be captured by a single number, but requires a complete risk profile.\\

\FloatBarrier
\section{The major share of exports are produced domestically}~\\
Here we quantify the relative contribution of domestic production $o^i_c$ and imports $h^i_c$ to the exports of Ukraine. As the allocation step uses fixed, country- $c$ and product- $i$ dependent shares $\eta_{c,i}^\mathrm{exp}$, the
domestic production accounts for

\begin{equation}
    \tilde{o}^i_c = \frac{\eta_{c,i}^\mathrm{exp}o^i_c(t=1)}{\eta_{c,i}^\mathrm{exp}(o_c^i(t=1)+h_c^i(t=1))} = \frac{o^i_c(t=1)}{o_c^i(t=1)+h_c^i(t=1)} 
\end{equation}

of the exports and the imports for

\begin{equation}
    \tilde{h}^i_c = \frac{\eta_{c,i}^\mathrm{exp}h^i_c(t=1)}{\eta_{c,i}^\mathrm{exp}(o_c^i(t=1)+h_c^i(t=1))} = \frac{h^i_c(t=1)}{o_c^i(t=1)+h_c^i(t=1)}~. 
\end{equation}

\begin{figure*}[h]
    \centering
    \includegraphics[width=0.9\textwidth]{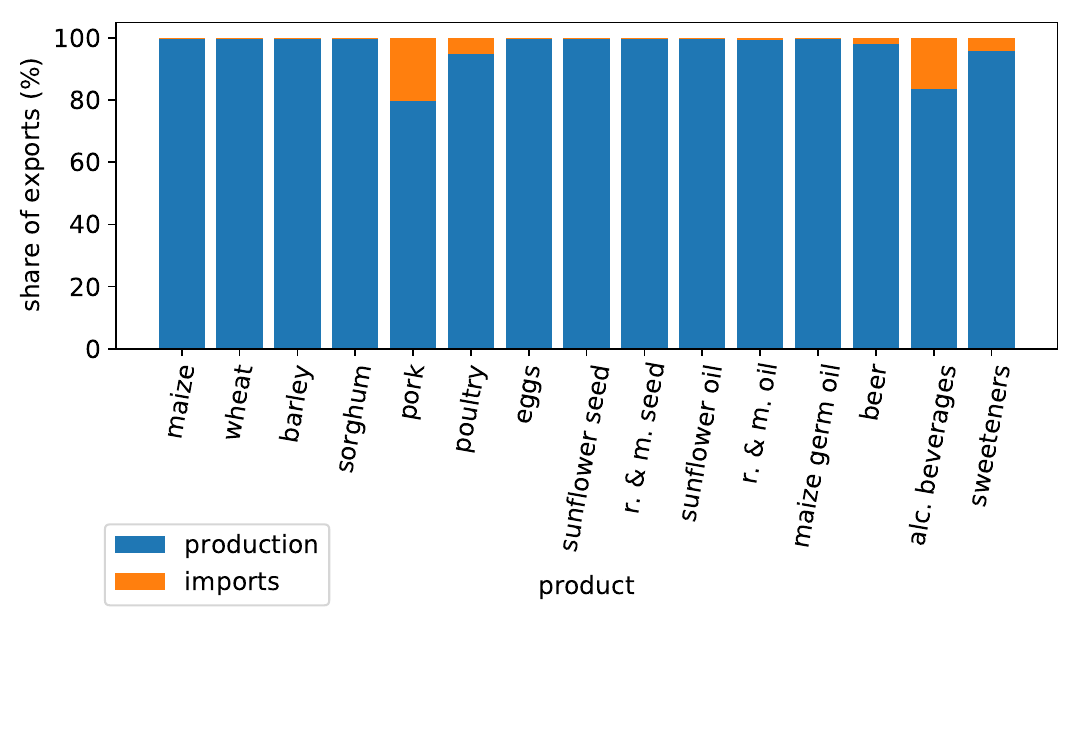}
    \caption{Share of domestic production (blue) and imports (orange) in Ukrainian exports. For most products the amount of reexported products is small.}
    \label{fig:reexports}
\end{figure*}

Figure~\ref{fig:reexports} shows how much production (blue) and imports (orange) contribute to the exports of agricultural products in the case of Ukraine. The Ukrainian exports of agricultural products consist largely of products produced in the country itself. Imported products contribute only little to the countries exports and therefore reexports are negligible. This means that the losses within the same layer are a good measure for the effects associated with trade, while the shocks across layers capture effects associated with production.

\section{Network Properties}~\\
Here we present the properties of the weighted, directed trade network $\tilde{T}^i$ of a product $i$, where link weights

\begin{equation}
    \tilde{T}'^{i}_{cd} = T^i_{cd}\eta_{d,i}^\mathrm{exp}x^i_d(t=0)
\end{equation}

are the exports of country $d$ to country $c$ in the first time step. We disregard links with a weight $\tilde{T}^i_{cd}<1\,\mathrm{t}$ and denote the weighted network after application of the threshold $\tilde{T}$.
We define the adjacency matrix $A^i$ for the network layer $i$ as

\begin{equation}
    A^i_{cd} = \left\{ \begin{array}{ll} 
         1  & \mathrm{if~~}\tilde{T}^i_{cd}>0  \\
         0  &  \mathrm{else}   \end{array} \right.~.
\end{equation}

We denote the total number of nodes $N$ and the total number of links $L$. We calculate the average number of trading partners, i.e. the average degree $\langle k^i \rangle$ on the $i$th layer as

\begin{equation}
    \langle k^i \rangle = \frac{1}{N}\sum_{c,d}\tilde{A}^i_{c,d}~.
\end{equation}

The out-degree $k^{i,\mathrm{out}}_c$ of country $c$ on layer $i$, i.e. the number of countries to which country $c$ exports product $i$, is

\begin{equation}
   k^{i,\mathrm{out}}_c = \sum_{d}\tilde{A}^i_{d,c}~.
\end{equation}

The average trade volume, i.e. the average node strength $\langle s^i \rangle$ on layer $i$ is

\begin{equation}
    \langle s^i \rangle = \frac{1}{N}\sum_{c,d}\tilde{T}^i_{c,d}~,
\end{equation}

while the out-strength of a single country $c$ on layer $i$ is

\begin{equation}
    s^{i,\mathrm{out}}_c = \sum_{d}\tilde{T}^i_{d,c}~.
\end{equation}

We denote the number of nodes in the largest strongly (weakly) connected component $N_\mathrm{SCC}$ ($N_\mathrm{WCC}$). In a strongly connected component each node can be reached from every other node following the directed links of the network. In a weakly connected component each node can be reached from every other node ignoring the direction of the links. Finally we report the Herfindhal index,

\begin{equation}
    H^i = \sum_{c}\left(\frac{s^{i,\mathrm{out}}_c}{\sum_{d}s^{i,\mathrm{out}}_d}\right)^2~,
\end{equation}

as a measure of market concentration and give the second moment of the out-degree distribution

\begin{equation}
    \langle (k^{i,\mathrm{out}})^2 \rangle = \frac{1}{N}\sum_{c} (k^{i,\mathrm{out}}_c)^2~,
\end{equation}

and second moment of the out-strength distribution

\begin{equation}
   \langle (s^{i,\mathrm{out}})^2 \rangle = \frac{1}{N}\sum_{c} (s^{i,\mathrm{out}}_c)^2~.
\end{equation}

Tables~\ref{tab:network_properties-1} and~\ref{tab:network_properties-2} summarize these quantities for the layers studied in the main text. Ukraine's high ranking in terms of node-strength and to a lesser extent also degree underlines the countries role as a hub in the trade network. The second moment of degree and strength distribution as well as the Herfindhal index show a substantial amount of heterogeneity in the network.

\begin{table}[h]
\centering
\begin{tabular}{lccccccccccccc}
\toprule
product $i$ &    $N^i$ & $N^i_\mathrm{SCC}$ & $N^i_\mathrm{WCC}$ & $L^i$ &  $\langle k^i\rangle$ & $\langle (k^{i,\mathrm{out}})^2\rangle$ & {$k^{i,\mathrm{out}}_\mathrm{UKR}$} & {$r^{i,k_{out}}_\mathrm{UKR}$} \\
\midrule
maize                   &  192 &   108 &   182 &   8,025 &  41.8 & 5,285.0   & 178 & 1  \\
wheat                   &  192 &   111 &   181 &  12,487 &  65.0 & 10,331.4  & 180 & 1  \\
barley                  &  192 &    67 &   174 &   5,759 &  30.0 & 4,119.7   & 105 & 34 \\
sorghum                 &  192 &    33 &   110 &   1,064 &  5.5  & 283.3    & 70	 & 5  \\
{pork}&  192 &    96 &   172 &   5,940 &  30.9 & 3,993.5   & 77	 & 40 \\
{poultry}           &  192 &   105 &   182 &   7,864 &  41.0 & 5,889.3   & 113 & 42 \\
eggs                    &  192 &    89 &   165 &   3,962 &  20.6 & 1,915.2   & 123 & 11 \\
sunflower seed          &  192 &    51 &   136 &   1,922 &  10.0 & 729.0    & 67	 & 12 \\
r. \& m. seed           &  192 &    50 &   151 &   2,815 &  14.7 & 1,386.6   & 146 & 1  \\
sunflower oil           &  192 &    59 &   156 &   4,105 &  21.4 & 2,610.6   & 153 & 1  \\
r. \& m. oil            &  192 &    50 &   140 &   2,868 &  14.9 & 1,339.5   & 77	 & 21 \\
maize germ oil          &  192 &    30 &   149 &   1,459 &   7.6 & 660.5    & 60	 & 12 \\
beer                    &  192 &   132 &   178 &   8,446 &  44.0 & 5,902.3   & 112 & 45 \\
alc. beverages          &  192 &   110 &   169 &   8,264 &  43.0 & 5,854.1   & 154 & 15 \\
sweeteners              &  192 &    69 &   172 &   7,145 &  37.2 & 5,625.4   & 149 & 29 \\
\bottomrule
\end{tabular}
\caption{Basic properties of the trade network layers for different products $i$. We report the total number of nodes $N^i$, the number of nodes in the largest strongly (weakly) connected component $N^i_\mathrm{SCC}$ ($N^i_\mathrm{WCC}$), the number of links $L^i$, the average degree $\langle k^i \rangle$, the second moment of the out-degree distribution $\langle (k^{i,\mathrm{out}})^2 \rangle$, the out-degree of Ukraine $k^{i,\mathrm{out}}_\mathrm{UKR}$ and its rank based on out-degree $r^{i,k_{out}}_\mathrm{UKR}$. We abbreviate {rapeseed and mustardseed} as {r.\& m.}.}\label{tab:network_properties-1}
\end{table}

\begin{table}[h]
\centering
\begin{tabular}{lccccc}
\toprule
product $i$ &    $\langle s^i \rangle$ & $\langle (s^{i,\mathrm{out}})^2 \rangle$ & $s^{i,\mathrm{out}}_\mathrm{UKR}$ & $r^{i,s_\mathrm{out}}_\mathrm{UKR}$ &    $H^i$\\
\midrule
maize                    &   632,290& 10,316,135,664,288  & 16,373,951 & 4	& 0.13  \\
wheat                    &  893,654 & 13,323,230,949,095  & 8,095,568  & 7	& 0.09  \\
barley                   & 196,319  & 82,4759,396,733     & 2,264,744  & 5	& 0.11  \\
sorghum                  &  32,930  & 63,618,431,756      & 180,832   & 5	& 0.31  \\
{pork}                   &  65,082  & 74,281,685,739      & 7,063	 & 39   & 0.09  \\
{pork}                   &  68,936  & 155,113,010,787     & 128,379	 & 16   & 0.17  \\
eggs                     & 11,341   & 2,174,190,579       & 45,226    & 13   & 0.09  \\
sunflower seed           & 21,994   & 15,465,562,303      & 63,150    & 13   & 0.17  \\
r. \& m. seed            & 100,050  & 379,688,828,438     & 2,509,323  & 3	& 0.20  \\
sunflower oil            & 36,062   & 63,805,941,791      & 3,138,201  & 1	& 0.26  \\
r. \& m. oil             & 27,612   & 25,526,095,706      & 40,437    & 15   & 0.17  \\
maize germ oil           &  2,849   & 705,027,575	     & 4,910     & 13   & 0.45  \\
beer                     & 58,945   & 53,756,667,087      & 235,260   & 11   & 0.08  \\
alc. beverages           & 15,507   & 4,354,294,892	     & 76,157    & 11   & 0.09  \\
sweeteners               & 40,727   & 50,585,702,025      & 61,363	 & 20   & 0.16  \\
\bottomrule
\end{tabular} 
\caption{Basic properties of the trade network layers for different products $i$ (continued). We report the average node strength $\langle s^i\rangle$, the second moment of the out-strength distribution $\langle (s^{i,\mathrm{out}})^2 \rangle$, the out strength of Ukraine $s^{i,\mathrm{out}}_\mathrm{UKR}$, Ukraines' ranking with respect to out-strength $r^{i,s_\mathrm{out}}_\mathrm{UKR}$ and the Herfindhal index $H^i$. The average strength is reported in tonnes, except for eggs, where we report counts. We abbreviate {rapeseed and mustardseed} as {r.\& m.}.}\label{tab:network_properties-2}
\end{table}

\section{Relative loss}~\\
In the following we list the value of the relative loss that occurs in different world regions in different shock scenarios.

\begin{table}[h]
\centering
\begin{tabular}{lrrrrrrrrr}
\toprule
{} &  maize &  wheat &  barley &  sorghum &  snflw. seed &  snflw. oil &  r.\&.m seed &  rape oil &  maize g. oil \\
\midrule
Africa N  &   17.1 &   24.7 &     6.8 &      0.2 &          1.8 &        48.3 &        5.0 &       9.6 &           1.2 \\
Africa S  &    1.9 &    7.8 &     0.0 &      0.1 &          0.0 &         4.5 &        0.8 &       2.0 &           1.6 \\
America S &    1.7 &    0.1 &     0.0 &      0.2 &          0.0 &         1.1 &        2.0 &       2.4 &           2.2 \\
America N &    0.1 &    0.0 &     0.0 &      0.1 &          0.6 &         6.7 &        0.0 &       0.1 &           0.1 \\
Asia C    &    2.2 &    1.1 &     0.0 &      0.0 &          0.1 &         5.1 &        0.2 &       0.3 &           1.8 \\
Asia E    &    1.0 &    0.3 &     1.7 &      0.6 &          0.0 &        48.8 &        0.1 &       1.5 &           1.6 \\
Asia S    &    5.9 &    0.9 &     0.1 &      0.0 &          0.5 &        67.8 &        1.8 &       2.2 &           6.5 \\
Asia SE   &    2.6 &    4.4 &     0.1 &      0.2 &          0.0 &         7.1 &        2.4 &       6.4 &           1.9 \\
Asia W    &   22.2 &    2.6 &    10.4 &      1.9 &          5.2 &        27.0 &       14.9 &      15.8 &          10.8 \\
Australia &    9.3 &    0.2 &     0.2 &      0.0 &          0.0 &         8.5 &        0.0 &       1.8 &           0.7 \\
Europe E  &    5.4 &    0.9 &     0.5 &      1.5 &          0.2 &         2.3 &        6.8 &       9.0 &           5.8 \\
Europe N  &   39.1 &    0.3 &     0.1 &      1.5 &          1.8 &        38.2 &        4.2 &       6.1 &           6.8 \\
Eurpe S   &   30.1 &    0.3 &     0.1 &     38.8 &          0.3 &        12.5 &       17.1 &      18.8 &          23.0 \\
Europe W  &   15.7 &    0.1 &     0.1 &      0.7 &          0.6 &        10.3 &       22.1 &      21.1 &           4.8 \\
\bottomrule
\end{tabular}
\caption{Relative loss of different food products in different world regions after a shock to the complete agricultural production of Ukraine. We abbreviate sunflower as snflw. and use r.\&m. seed to refer to rapeseed and mustardseed.These numbers correspond to fig.\,4 in the main text.}
\end{table}

\begin{table}[h]
\centering
\begin{tabular}{lrrrrrr}
\toprule
{} &  pork &  poultry &  eggs &  beer &  alc. beverages &  sweeteners \\
\midrule
Africa N  &       6.6 &          12.4 &  12.2 &   3.2 &             0.9 &         5.7 \\
Africa S  &       1.2 &           1.3 &   0.6 &   1.4 &             0.8 &         0.4 \\
America S &       0.9 &           1.0 &   1.5 &   1.0 &             0.6 &         0.5 \\
America N &       0.1 &           0.1 &   0.1 &   0.4 &             0.6 &         0.2 \\
Asia C    &       1.3 &           8.0 &   0.8 &   1.0 &             2.4 &        28.9 \\
Asia E    &       0.4 &           0.5 &   0.5 &   0.9 &             0.3 &         1.4 \\
Asia S    &       0.1 &           5.8 &   2.4 &   0.1 &             0.1 &         0.2 \\
Asia SE   &       0.8 &           1.1 &   1.2 &   0.6 &             0.6 &         0.5 \\
Asia W    &       7.0 &           6.8 &  10.0 &   3.8 &             3.5 &         2.4 \\
Australia &       0.9 &           0.4 &   0.4 &   1.1 &             2.2 &         0.3 \\
Europe E  &       2.9 &           2.9 &   2.1 &   2.9 &             4.3 &         3.2 \\
Europe N  &       3.2 &           3.4 &   3.4 &   0.9 &             4.3 &         3.3 \\
Eurpe S   &      12.9 &          17.2 &  17.2 &   4.6 &             1.0 &        10.7 \\
Europe W  &       4.9 &           5.0 &   5.6 &   4.7 &             2.3 &         3.5 \\
\bottomrule
\end{tabular}
\caption{Relative loss of different food products in different world regions after a shock to the complete agricultural production of Ukraine. These numbers correspond to fig. 4 in the main text.}
\end{table}

\begin{table}[h]
\centering
\begin{tabular}{lrrrrrrrr}
\toprule
{} &  maize &  pork &  poultry &  eggs &  maize g. oil &  beer &  alc. beverages &  sweeteners \\
\midrule
Africa N  &   17.1 &       2.5 &           4.9 &   4.9 &           1.2 &   0.2 &             0.7 &         5.5 \\
Africa S  &    1.9 &       1.0 &           0.9 &   0.4 &           1.6 &   1.3 &             0.3 &         0.3 \\
America S &    1.7 &       0.8 &           0.9 &   1.4 &           2.2 &   0.9 &             0.6 &         0.4 \\
America N &    0.1 &       0.1 &           0.0 &   0.0 &           0.1 &   0.4 &             0.4 &         0.0 \\
Asia C    &    2.2 &       0.7 &           2.5 &   0.3 &           1.8 &   0.3 &             0.7 &        11.4 \\
Asia E    &    1.0 &       0.3 &           0.5 &   0.4 &           1.6 &   0.8 &             0.3 &         1.2 \\
Asia S    &    5.9 &       0.1 &           5.6 &   2.3 &           6.5 &   0.0 &             0.0 &         0.1 \\
Asia SE   &    2.6 &       0.6 &           0.8 &   0.8 &           1.9 &   0.4 &             0.2 &         0.2 \\
Asia W    &   22.2 &       5.1 &           3.9 &   5.5 &          10.8 &   2.1 &             2.4 &         1.1 \\
Australia &    9.3 &       0.6 &           0.3 &   0.3 &           0.7 &   0.7 &             1.7 &         0.2 \\
Europe E  &    5.4 &       1.7 &           1.4 &   1.1 &           5.8 &   1.9 &             1.4 &         1.3 \\
Europe N  &   39.1 &       2.1 &           2.6 &   2.6 &           6.8 &   0.7 &             3.6 &         2.3 \\
Eurpe S   &   30.1 &      11.4 &          15.8 &  15.8 &          23.0 &   4.5 &             0.7 &        10.0 \\
Europe W  &   15.7 &       3.4 &           3.9 &   4.4 &           4.8 &   4.6 &             1.8 &         3.2 \\
\bottomrule
\end{tabular}
\caption{Relative loss in \% of different food products in different world regions after a shock to the maize production of Ukraine. These numbers correspond to left half of each cell in the upper part of fig. 5 in the main text.}
\end{table}

\begin{table}[h]
\centering
\begin{tabular}{lrrrrrrrr}
\toprule
{} &  maize &  pork &  poultry &  eggs &  maize g. oil &  beer &  alc. beverages &  sweeteners \\
\midrule
Africa N  &   17.1 &       0.0 &           0.2 &   0.1 &           0.1 &   0.0 &             0.1 &         0.1 \\
Africa S  &    1.9 &       0.0 &           0.1 &   0.2 &           0.0 &   0.0 &             0.1 &         0.1 \\
America S &    1.7 &       0.0 &           0.0 &   0.0 &           0.0 &   0.0 &             0.0 &         0.1 \\
America N &    0.1 &       0.0 &           0.0 &   0.0 &           0.0 &   0.0 &             0.1 &         0.1 \\
Asia C    &    2.2 &       0.0 &           7.2 &   0.0 &           1.4 &   0.4 &             1.5 &        28.3 \\
Asia E    &    1.0 &       0.0 &           0.0 &   0.0 &           0.0 &   0.0 &             0.0 &         0.0 \\
Asia S    &    5.9 &       0.0 &           0.0 &   0.0 &           1.2 &   0.0 &             0.0 &         0.0 \\
Asia SE   &    2.6 &       0.0 &           0.0 &   0.0 &           0.1 &   0.0 &             0.0 &         0.0 \\
Asia W    &   22.2 &       0.0 &           0.5 &   1.6 &           1.1 &   0.1 &             0.5 &         1.2 \\
Australia &    9.3 &       0.0 &           0.0 &   0.0 &           0.1 &   0.0 &             0.2 &         0.1 \\
Europe E  &    5.4 &       0.1 &           0.6 &   0.1 &           4.6 &   1.3 &             2.8 &         1.7 \\
Europe N  &   39.1 &       0.0 &           0.0 &   0.0 &           0.0 &   0.1 &             0.1 &         0.3 \\
Eurpe S   &   30.1 &       0.0 &           0.0 &   0.0 &           0.0 &   0.0 &             0.1 &         0.0 \\
Europe W  &   15.7 &       0.0 &           0.0 &   0.0 &           0.0 &   0.0 &             0.1 &         0.1 \\
\bottomrule
\end{tabular}
\caption{Relative loss in \% of different food products in different world regions after a shock to the respective product. These numbers correspond to the right half of each cell in the upper part of fig. 5 in the main text.}
\end{table}

\begin{table}[h]
\centering
\begin{tabular}{lrrr}
\toprule
{} &  snflw. seed &  snflw. oil &  snflw. cake \\
\midrule
Africa N  &          1.8 &        48.2 &         42.9 \\
Africa S  &          0.0 &         4.5 &          2.1 \\
America S &          0.0 &         1.1 &          1.2 \\
America N &          0.6 &         6.7 &          1.1 \\
Asia C    &          0.1 &         5.1 &          0.2 \\
Asia E    &          0.0 &        48.7 &          0.0 \\
Asia S    &          0.5 &        67.7 &          1.2 \\
Asia SE   &          0.0 &         7.1 &          0.4 \\
Asia W    &          5.2 &        27.0 &         18.8 \\
Australia &          0.0 &         8.5 &          2.5 \\
Europe E  &          0.2 &         2.3 &         15.9 \\
Europe N  &          1.8 &        38.2 &         39.1 \\
Eurpe S   &          0.3 &        12.5 &         15.3 \\
Europe W  &          0.6 &        10.3 &         28.1 \\
\bottomrule
\end{tabular}

\caption{Relative loss in \% of different food products in different world regions after a shock to the sunflower seed production of Ukraine. We use the abbreviation snflw. for sunflower. These numbers correspond to the left half of each cell in the lower part of fig. 5 in the main text.}
\end{table}

\begin{table}
\centering
\begin{tabular}{lrrr}
\toprule
{} &  snflw. seed &  snflw. oil &  snflw. cake \\
\midrule
Africa N  &          1.8 &        48.1 &         42.4 \\
Africa S  &          0.0 &         4.5 &          2.1 \\
America S &          0.0 &         1.1 &          1.2 \\
America N &          0.6 &         6.3 &          0.6 \\
Asia C    &          0.1 &         5.0 &          0.1 \\
Asia E    &          0.0 &        48.7 &          0.0 \\
Asia S    &          0.5 &        67.7 &          0.8 \\
Asia SE   &          0.0 &         7.1 &          0.4 \\
Asia W    &          5.2 &        24.8 &         15.9 \\
Australia &          0.0 &         8.5 &          2.5 \\
Europe E  &          0.2 &         2.1 &         15.8 \\
Europe N  &          1.8 &        38.1 &         39.1 \\
Eurpe S   &          0.3 &        12.3 &         15.2 \\
Europe W  &          0.6 &         9.9 &         27.9 \\
\bottomrule
\end{tabular}
\caption{Relative loss of different food products in different world regions after a shock to the respective product. These numbers correspond to the right half of each cell in the lower part of fig. 5 in the main text.}
\end{table}

\FloatBarrier
\section{Sensitivity Analysis}

\begin{figure}[p]
    \centering
    \includegraphics{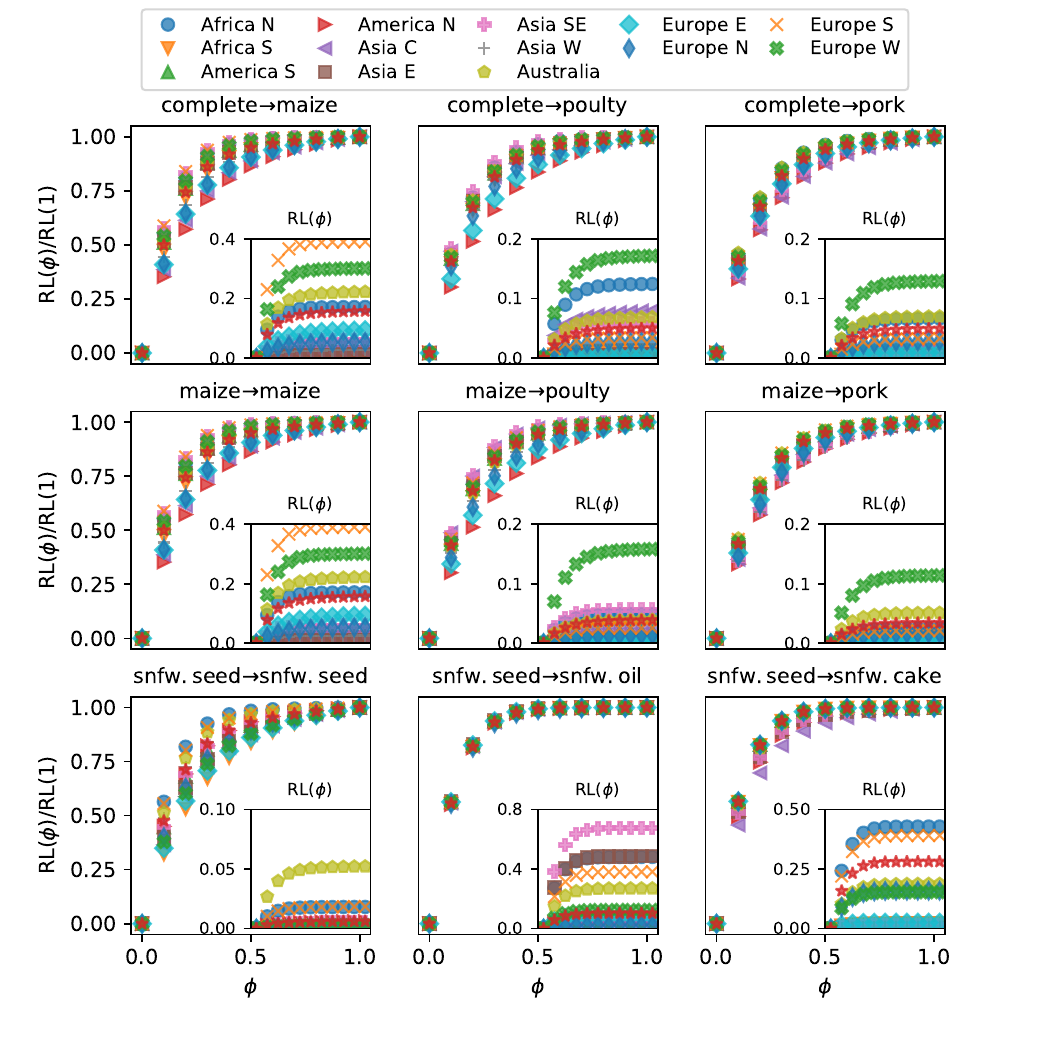}
    \caption{The relative loss $\mathrm{RL}(\phi)$ in different world regions after a shock to the production of selected food and agricultural products in Ukraine is displayed as a function of the shock size $\phi$. For each region we show the relative loss normalized to the relative loss for a maximal shock, $\mathrm{RL}(\phi=1)$ and present the unnormalized relative loss in the inset. The notation $i\to j$ means that product $i$ was shocked and losses for product $j$ are shown. A simultaneous shock to all products is called \textit{complete}. We use the shorthand \text{snfw.} for sunflower. The relative loss increases non-linearly with the shock size, $\phi$, and saturates for large values of $\phi$.}
    \label{fig:sensitivity}
\end{figure}

Figure~\ref{fig:sensitivity} shows how the losses of selected food products if we relax the worst case assumption of a complete loss of production in Ukraine and vary the shock size $\phi$. The shock size $\phi$ describes the fraction of production output that is lost (see eq.~9 in the main text). To highlight common patterns, the relative loss in each region is normalized to the relative loss that occurs for a complete shock, $\phi=1$. The unnormalized relative loss is shown in the inset. We introduce the notation $i\to j$ to denote that we show the losses of product $j$ after a shock to product $i$. A simultaneous shock to the production of all food products in Ukraine is called \textit{complete}.\\
We find that losses grow non-linearly with the shock size. The growth is rapid at first but saturation is observed at greater shock sizes. This non-linear behavior is tightly linked to the way the production of certain crops is modelled: Sowing grains and oil-seeds requires a certain amount of the same crop, thereby creating a feedback between available amount at time $t$ and available amount at time $t+1$.\\
As countries do not change the fraction of the available amount that they use for production, export and consumption, they cannot counter this trend by allocating more grain to the production of grain, e.g. on the expense of exports.

\FloatBarrier
\section{Parameters and Variables}
The following table~\ref{tab:params} lists all parameters and variables in the model and indicates whether they are static or evolve with time. It also indicates whether they are directly taken from data, calculated on the basis of data before the simulation (calibrated) or an assumptions.

\begin{table}[h]
    \centering
    \begin{tabular}{c|l|c|c}
    \toprule
    Parameter/Variable      &   Meaning     &   Origin & Static/Dynamic \\ \hline
    $\alpha_{c,i}^k$        & amount of output product $i$ produced by      &   calibrated  & static \\
                            & process $k$ in country $c$ per unit of input  &               &        \\
    $\beta_{c,i}^k$         & amount of output product $i$ produced         &  calibrated   & static \\
                            & by process $k$ in country $c$ if no input is available    &   &        \\
    $\mathcal{C}$           &  set of all countries                         &   data  & static \\
    $e_c^i(t)$              & the amount of product $i$ that                &   variable    & dynamic \\
                            & country $c$ exports                           &               &         \\
    $\eta_{c,i}^{prod}$     &  fraction of product $i$ country $c$          &  calibrated   & static  \\
                            &  allocates as input to production             &               &         \\
    $\eta_{c,i}^{exp}$      &  fraction of product $i$ country $c$          &  calibrated   & static  \\
                            & allocates as exports                          &               &         \\
    $\eta_{c,i}^{food}$     & fraction of product $i$ country $c$           &  calibrated   & static  \\
                            & allocates to consumption as food              &               &         \\
    $\eta_{c,i}^{else}$     & fraction of product $i$ country $c$           &  calibrated   & static  \\
                            & allocates to any other use                    &               &         \\
    $f^{i}_{c,k}(p_c)$      & production function describing how much of    &  calibrated   & static  \\
                            & output product $i$ can be                     &               & \\
                            & obtained in country $c$ by process type $k$   &               &\\
                            & given the available input $p_c(t)$            &               &\\
    $\mathcal{F}$           &  set of all products                          &  data         & static \\
    $h^i_c(t)$              & the imports of product $i$ in country $c$     &  variable     & dynamic\\
                            & from all trade partners                       &               & \\
    $\mathcal{I}_{c,k}$     & set of input products that process $k$        & calibrated    & static \\
                            & in country $c$ uses                           &               & \\
    $v_{c,i}^k$             & fraction of the production input $p^i_c$ that & calibrated    & static \\
                            & country $c$ assigns to process $k$.           &               &\\
    $o_c^i(t)$              & amount of product $i$ supplied by all processes in country $c$& variable & dynamic\\
    $\mathcal{O}_{c,k}$     & set of output products that process $k$ in country $c$ supplies & calibrated & static\\
    $p_c^i$                 & the amount of product $i$ that country $c$    &  variable      & dynamic\\
                            & uses  as input to production                  &                & \\
    $\phi$                  & relative shock size                            & assumption     & static \\
    $k_c^i(t)$              & the amount of product $i$ that country $c$     & variable       & dynamic\\
                            & uses for consumption as food                   &                &\\
    $r_c^i(t)$              & the amount of product $i$ that country $c$     & variable       & dynamic\\
                            & uses for any other purpose                     &                &\\
$\mathrm{RL}_{cd}^{ij}(t)$  & the losses of product $i$ in country $c$ after & output        & dynamic \\
                            & a shock of product $j$ in country $d$          &               &\\
    $S_{(c,k),(d,i)}$       & supply table: amount of product $i$ created in country $d$ &  data   & static\\
                            & by process $k$ of country $c$ & &\\
    $t$                     & a specific iteration (model time)             &   parameter    & dynamic  \\
    $t_\mathrm{end}$                  & number of iterations in both shocked          &   assumption   & static \\
                            & and baseline scenario (model time)            &                & \\
    $T^i_{cd}$              & share of the exports $e_d^i$ of product $i$ from & calibrated  & static\\
                            & country $d$ that are directed towards country $c$  &           & \\
    $U_{(d,i),(c,k)}$       & use table: amount of product $i$ created in country $d$ & data & static\\
                            & by process $k$ of country $c$ & &\\
    $x_c^i(t)$              & the total amount of product $i$ available in  &   variable    & dynamic \\ 
                            & country $c$ for any purpose & &\\
    $\tilde{x}_c^i$         & amount of product $i$ in country $c$ at           & calibrated & static \\
                            & the beginning of the first iteration              &            &\\
    $Y_{(d,i),(c,k)}$       & demand: amount of product $i$ from country $d$    & data & static \\ 
                            & needed to satisfy demand of type $k$ in country $c$ &          &\\
    $\underline{~}$         & indicates a quantity in the baseline scenario     & --- & --- \\
    \bottomrule
    \end{tabular}
    \caption{All symbols used in the model, ordered alphabetically.}
    \label{tab:params}
\end{table}

\FloatBarrier
\section{World Regions}
The following tables (tab.~\ref{tab:african-regions} to tab.~\ref{tab:regions-in-australia-and-others}) provide a mapping between the world regions used in the main text and the countries and regions of the FABIO multi-regional input--output model~\cite{Bruckner_2019-10}.

\begin{table}[h]
\centering
\begin{tabular}[t]{l|l}
\toprule
region &                                   country \\ \hline
Africa N &                                Algeria \\
Africa N &                                  Egypt \\
Africa N &                                  Libya \\
Africa N &                                Morocco \\
Africa N &                                  Sudan \\
Africa N &                                Tunisia \\
Africa S &                                 Angola \\
Africa S &                                  Benin \\
Africa S &                               Botswana \\
Africa S &                           Burkina Faso \\
Africa S &                                Burundi \\
Africa S &                             Cabo Verde \\
Africa S &                               Cameroon \\
Africa S &               Central African Republic \\
Africa S &                                   Chad \\
Africa S &                                  Congo \\
Africa S &                          Côte d'Ivoire \\
Africa S &                  Dem. Rep. of the Congo \\
Africa S &                               Djibouti \\
Africa S &                                Eritrea \\
Africa S &                               Eswatini \\
Africa S &                               Ethiopia \\
Africa S &                                  Gabon \\
Africa S &                                 Gambia \\
Africa S &                                  Ghana \\
Africa S &                                 Guinea \\
Africa S &                          Guinea-Bissau \\
\bottomrule
\end{tabular}
\hspace{0.5cm}
\begin{tabular}[t]{l|l}
\toprule
region   &                              country  \\ \hline
Africa S &                                  Kenya \\
Africa S &                                Lesotho \\
Africa S &                                Liberia \\
Africa S &                             Madagascar \\
Africa S &                                 Malawi \\
Africa S &                                   Mali \\
Africa S &                             Mauritania \\
Africa S &                              Mauritius \\
Africa S &                             Mozambique \\
Africa S &                                Namibia \\
Africa S &                                  Niger \\
Africa S &                                Nigeria \\
Africa S &                                 Rwanda \\
Africa S &                  Sao Tome and Principe \\
Africa S &                                Senegal \\
Africa S &                           Sierra Leone \\
Africa S &                                Somalia \\
Africa S &                           South Africa \\
Africa S &                            South Sudan \\
Africa S &                                   Togo \\
Africa S &                                 Uganda \\
Africa S &            United Republic of Tanzania \\
Africa S &                                 Zambia \\
Africa S &                               Zimbabwe \\
\bottomrule
\end{tabular}
\caption{Countries and  world regions of Africa}\label{tab:african-regions}
\end{table}

\begin{table}[h]
\centering
\begin{tabular}[t]{l|l}
\toprule
region    &                 country \\ \hline
America S &                    Antigua and Barbuda \\
America S &                              Argentina \\
America S &                                Bahamas \\
America S &                               Barbados \\
America S &                                 Belize \\
America S &       Bolivia (Plurinational State of) \\
America S &                                 Brazil \\
America S &                                  Chile \\
America S &                               Colombia \\
America S &                             Costa Rica \\
America S &                                   Cuba \\
America S &                               Dominica \\
America S &                     Dominican Republic \\
America S &                                Ecuador \\
America S &                            El Salvador \\
America S &                                Grenada \\
America S &                              Guatemala \\
America S &                                 Guyana \\
America S &                                  Haiti \\
America S &                               Honduras \\
America S &                                Jamaica \\
America S &                                 Mexico \\
America S &                              Nicaragua \\
America S &                                 Panama \\
America S &                               Paraguay \\
America S &                                   Peru \\
America S &                            Puerto Rico \\
America S &                  Saint Kitts and Nevis \\
America S &                            Saint Lucia \\
America S &             Saint Vincent and the Grenadines \\
America S &                               Suriname \\
America S &                    Trinidad and Tobago \\
America S &                                Uruguay \\
America S &     Venezuela (Bolivarian Republic of) \\
America N &                                 Canada \\
America N &               United States of America \\
\bottomrule
\end{tabular}
\caption{Countries and world regions in the Americas}\label{tab:regions-in-americas}
\end{table}

\begin{table}[h]
\centering
\begin{tabular}[t]{l|l}
\toprule
region & country \\ \hline
Asia C &                             Kazakhstan \\
Asia C &                             Kyrgyzstan \\
Asia C &                             Tajikistan \\
Asia C &                           Turkmenistan \\
Asia C &                             Uzbekistan \\
Asia E &                   China, Hong Kong SAR \\
Asia E &                       China, Macao SAR \\
Asia E &              China, Taiwan Province of \\
Asia E &                        China, mainland \\
Asia E &            Dem. People's Rep. of Korea \\
Asia E &                                  Japan \\
Asia E &                               Mongolia \\
Asia E &                      Republic of Korea \\
Asia S &                            Afghanistan \\
Asia S &                             Bangladesh \\
Asia S &                                  India \\
Asia S &             Iran (Islamic Republic of) \\
Asia S &                               Maldives \\
Asia S &                                  Nepal \\
Asia S &                               Pakistan \\
Asia S &                              Sri Lanka \\
Asia SE &                      Brunei Darussalam \\
Asia SE &                               Cambodia \\
Asia SE &                              Indonesia \\
Asia SE &                Lao People's Dem. Rep. \\
\bottomrule
\end{tabular}
\hspace{0.2cm}
\begin{tabular}[t]{l|l}
\toprule
region & country \\ \hline
Asia SE &                               Malaysia \\
Asia SE &                                Myanmar \\
Asia SE &                            Philippines \\
Asia SE &                              Singapore \\
Asia SE &                               Thailand \\
Asia SE &                            Timor-Leste \\
Asia SE &                               Viet Nam \\
Asia W &                                Armenia \\
Asia W &                             Azerbaijan \\
Asia W &                                Bahrain \\
Asia W &                                 Cyprus \\
Asia W &                                Georgia \\
Asia W &                                   Iraq \\
Asia W &                                 Israel \\
Asia W &                                 Jordan \\
Asia W &                                 Kuwait \\
Asia W &                                Lebanon \\
Asia W &                                   Oman \\
Asia W &                                  Qatar \\
Asia W &                           Saudi Arabia \\
Asia W &                   Syrian Arab Republic \\
Asia W &                                 Turkey \\
Asia W &                   United Arab Emirates \\
Asia W &                                  Yemen \\
\bottomrule
\end{tabular}
\caption{Countries and world regions of Asia}\label{regions-in-asia}
\end{table}

\begin{table}[H]
\centering
\begin{tabular}[t]{l|l}
\toprule
region & country \\ \hline
Europe E &                                Belarus \\
Europe E &                               Bulgaria \\
Europe E &                         Czech Republic \\
Europe E &                                Hungary \\
Europe E &                                 Poland \\
Europe E &                    Republic of Moldova \\
Europe E &                                Romania \\
Europe E &                     Russian Federation \\
Europe E &                               Slovakia \\
Europe E &                                Ukraine \\
Europe N &                                Denmark \\
Europe N &                                Estonia \\
Europe N &                                Finland \\
Europe N &                                Iceland \\
Europe N &                                Ireland \\
Europe N &                                 Latvia \\
Europe N &                              Lithuania \\
Europe N &                                 Norway \\
Europe N &                                 Sweden \\
Europe N &                         United Kingdom \\
Europe S &                                Albania \\
Europe S &                 Bosnia and Herzegovina \\
Europe S &                                Croatia \\
Europe S &                                 Greece \\
Europe S &                                  Italy \\
Europe S &                                  Malta \\
Europe S &                             Montenegro \\
Europe S &                        North Macedonia \\
Europe S &                               Portugal \\
Europe S &                                 Serbia \\
Europe S &                               Slovenia \\
Europe S &                                  Spain \\
Europe W &                                Austria \\
Europe W &                                Belgium \\
Europe W &                                 France \\
Europe W &                                Germany \\
Europe W &                             Luxembourg \\
Europe W &                            Netherlands \\
Europe W &                            Switzerland \\
\bottomrule
\end{tabular}
\caption{Countries and world regions of Europe}\label{tab:regions-in-europe}
\end{table}

\begin{table}[H]
\centering
\begin{tabular}{l|l} 
\toprule
region & country \\ \hline
Australia                 &  Australia \\
Australia                 &  New Zealand \\
Melanesia &                                   Fiji \\
Melanesia &                          New Caledonia \\
Melanesia &                       Papua New Guinea \\
Melanesia &                        Solomon Islands \\
Melanesia &                                Vanuatu \\
Micronesia &                               Kiribati \\
Polynesia &                       French Polynesia \\
Polynesia &                                  Samoa \\
not classified &                     Belgium-Luxembourg \\
not classified &                         Czechoslovakia \\
not classified &                   Netherlands Antilles \\
not classified &                                    RoW \\
not classified &                  Serbia and Montenegro \\
not classified &                                   USSR \\
not classified &                           Yugoslav SFR \\
\bottomrule
\end{tabular}
\caption{Regions of Australia, New Zealand and Oceania and not classified}\label{tab:regions-in-australia-and-others}
\end{table}

\FloatBarrier
\section{Algorithm}
Here, we present pseudo-code for the algorithm in the shocked scenario in the case of a shock. If the evaluation of a production function $f_{c,i}^{k}$ for one output product $i$ takes asymptotically time $\mathcal{O}(\mathrm{T}_f)$, the worst case time complexity of this algorithm is 

\begin{equation}
    \mathcal{O}(|\mathcal{C}||\mathcal{F}|(|\mathcal{P}|\mathrm{T}_f + |\mathcal{C}|))~.
\end{equation}

Note that for all practical purposes, the worst case is not realized: As production processes usually provide a small number of output products the factor $|\mathcal{P}|$ could be replaced by a constant.
In the following, we denote the shocked countries as $\mathcal{SC}$ and the shocked products as $\mathcal{SP}$. All other notation is introduced in the methods section and summarized in table~\ref{tab:params}~.

\begin{algorithm}
\begin{algorithmic}[1]
\State  \# Initial Condition \#
\For{$(c,i) \in \mathcal{C}\times\mathcal{F}$}
   \State $e_c^i(0) \gets \eta_{c,i}^\mathrm{exp}  x_c^i(0)$
   \State $p_c^i(0) \gets \eta_{c,i}^\mathrm{prod} x_c^i(0)$
   \State $k_c^i(0) \gets \eta_{c,i}^\mathrm{food} x_c^i(0)$
   \State $r_c^i(0) \gets \eta_{c,i}^\mathrm{else} x_c^i(0)$
\EndFor
\State~
\For{$t \in \{0,\dots,\tau\}$}~\\
    \State \# Reset \#
    \If{$t=0$}
     \State   $x(t) \gets \tilde{x}$
    \Else
     \State $x(t) \gets 0$
    \EndIf
\State ~
\State    \# Production \#
    
    \For{$(c,k,i) \in \mathcal{C}\times\mathcal{P}\times\mathcal{F}$}
        \State $x_c^i(t) \gets x_c^i(t) +  f^k_{c,i}(p_c^1(t-1),\dots,p_c^{|\mathcal{F}|}(t-1))$
    \EndFor
\State~
\State   \# Shock \#
    \For{$(c,i) \in\mathcal{SC}\times\mathcal{SF}$}
      \State  $x_c^i(t) \gets 0$
    \EndFor
\State~
\State    \# Trade  \#
    \For{$(c,i) \in \mathcal{C}\times\mathcal{F}$}
      \State  $x_c^i(t) \gets x_c^i(t) + \sum_{d\in\mathcal{C}} T_{cd}^i e_d^i(t-1)$
    \EndFor
\State~
\State    \# Allocation \#
    
    \For{$(c,i) \in \mathcal{C}\times\mathcal{F}$}
   \State $e_c^i(t) \gets \eta_{c,i}^\mathrm{exp}  x_c^i(t)$
   \State $p_c^i(t) \gets \eta_{c,i}^\mathrm{prod} x_c^i(t)$
   \State $k_c^i(t) \gets \eta_{c,i}^\mathrm{food} x_c^i(t)$
   \State $r_c^i(t) \gets \eta_{c,i}^\mathrm{else} x_c^i(t)$
    \EndFor
\EndFor

\end{algorithmic}
\caption{Pseudo-code for the shocked scenario. The pseudo-code for the baseline scenario is easily obtained by omitting the lines 23-26.}
\end{algorithm}

\end{document}